\documentclass[preprint,aps,pra,amsmath,superscriptaddress]{revtex4-1}
\usepackage{amssymb}
\usepackage{mathrsfs}
\usepackage{graphicx}
\usepackage{subfigure}
\usepackage{float}
\usepackage[english]{babel}
\usepackage[utf8x]{inputenc}
\usepackage[T1]{fontenc}
\usepackage{color}
\usepackage{CJK}

\newlength{\figurewidth}
\begin{document}
\begin{CJK*}{UTF8}{}

\title{Theoretical evidence for the sensitivity of charge-rearrangement-enhanced x-ray ionization to molecular size}

\author{Yajiang Hao \CJKfamily{gbsn}(郝亚江)}
\email{haoyj@ustb.edu.cn}
\affiliation{Department of Physics, University of Science and Technology Beijing, Beijing 100083, P.~R.~China}
\author{Ludger Inhester}
\email{ludger.inhester@cfel.de}
\affiliation{Center for Free-Electron Laser Science, DESY, Notkestrasse 85, 22607 Hamburg, Germany}
\affiliation{The Hamburg Centre for Ultrafast Imaging, Luruper Chaussee 149, 22761 Hamburg, Germany}
\author{Sang-Kil Son \CJKfamily{mj}(손상길)}
\email{sangkil.son@cfel.de}
\affiliation{Center for Free-Electron Laser Science, DESY, Notkestrasse 85, 22607 Hamburg, Germany}
\affiliation{The Hamburg Centre for Ultrafast Imaging, Luruper Chaussee 149, 22761 Hamburg, Germany}
\author{Robin Santra}
\email{robin.santra@cfel.de}
\affiliation{Center for Free-Electron Laser Science, DESY, Notkestrasse 85, 22607 Hamburg, Germany}
\affiliation{The Hamburg Centre for Ultrafast Imaging, Luruper Chaussee 149, 22761 Hamburg, Germany}
\affiliation{Department of Physics, University of Hamburg, Jungiusstrasse 9, 20355 Hamburg, Germany}
\date{\today}

\begin{abstract}
It was recently discovered that molecular ionization at high x-ray intensity is enhanced, in comparison with that of isolated atoms, through a phenomenon called CREXIM (charge-rearrangement-enhanced x-ray ionization of molecules).
X-ray absorption selectively ionizes heavy atoms within molecules, triggering electron transfer from neighboring atoms to the heavy atom sites and enabling further ionization there.
The present theoretical study demonstrates that the CREXIM effect increases with the size of the molecule, as a consequence of increased intramolecular electron transfer from the larger molecular constituents attached to the heavy atoms.
We compare x-ray multiphoton ionization dynamics of xenon, iodomethane, and iodobenzene after interacting with an intense x-ray pulse.
Although their photoionization cross sections are similar, iodomethane and iodobenzene molecules are more ionized than xenon atoms.
Moreover, we predict that the average total charge of iodobenzene is much larger than that of iodomethane, because of the large number of electrons in the benzene ring.
The positive charges transferred from the iodine site to the benzene ring are redistributed such that the higher carbon charges are formed at the far end from the iodine site.
Our first-principles calculations provide fundamental insights into the interaction of molecules with x-ray free-electron laser (XFEL) pulses.
These insights need to be taken into account for interpreting and designing future XFEL experiments.
\end{abstract}
\maketitle
\end{CJK*}


X-ray free-electron lasers (XFELs) have opened new horizons for studying the structure and dynamics of matter with femtosecond temporal and atomic spatial resolution~\cite{Marangos11,Pellegrini16,Bostedt16}.
With the rapid advance in XFEL technology, the peak intensity already exceeds $10^{20}$~W/cm$^2$ in the hard x-ray regime~\cite{Mimura14}.
At extremely high x-ray intensity, it is crucial to understand the nonlinear interaction of atoms~\cite{Young10,Doumy11,Rudek12,Fukuzawa13,Rudek18}, molecules~\cite{Hoener10,Fang10,Erk13,Erk13a,Boll16,Motomura15,Erk14,Rudenko17}, and clusters~\cite{Murphy14,Tachibana15,Kumagai18} with x-ray fields.
The interaction between matter and intense x-ray pulses can be described by the multiphoton multiple ionization dynamics model.
An x-ray photon ionizes an atomic inner-shell electron, followed by relaxation processes.
During the intense x-ray pulse the photoionization and accompanying processes repeat sequentially.
This ionization model has been verified for atomic systems with a series of gas-phase XFEL experiments~\cite{Young10,Rudek12,Fukuzawa13}.

When a molecule is irradiated by x rays, they mainly interact with heavy atoms within the molecule, because their photoionization cross section is much higher than that of light atoms.
It had been widely believed that ionization dynamics of a molecule exposed to XFEL irradiation is similar to that of an isolated atom whose cross section is comparable to the molecule~\cite{Erk13,Erk13a,Motomura15,Boll16}. 
This expectation turned out to be valid in the low-fluence regime.
For example, the total molecular ionization of methylselenol (CH$_3$SeH) at low x-ray intensity, where single-photon absorption is dominant, was found to be similar to the atomic ionization of Kr under the same x-ray beam conditions~\cite{Erk13}.
In contrast, molecular ionization at high x-ray intensity, where multiphoton absorption becomes dominant, is fundamentally different from atomic ionization.
A recent XFEL experiment on iodomethane~\cite{Rudenko17} demonstrated that the total molecular charge is much higher than the sum of the atomic charges within the independent-atom model.
Theory~\cite{Inhester16,Rudenko17} explained that the molecular ionization enhancement is due to recurrent charge rearrangement upon each x-ray absorption event.
In the molecular case, an x-ray-irradiated heavy atom pulls in electrons from neighboring atoms and ejects the electrons from there, acting like a \emph{molecular black hole}~\cite{blackhole}.

This striking feature of molecular ionization enhancement at high x-ray intensity invokes immediate questions.
Is this effect more pronounced for larger molecular systems?
How much and how far does it affect neighboring atoms as a function of distance?
Understanding the dynamical behavior of heavy-atom-containing molecules is critical for serial femtosecond crystallography~\cite{Chapman11,Boutet12} and single-particle imaging~\cite{Aquila15,Yoon16,Fortmann-Grote17}, where ultrashort and ultraintense x-ray pulses are desirable.
In particular, it has been proposed to take advantage of severe radiation damage on heavy atoms for novel phasing methods~\cite{Son11e,Galli15b}.
Knowledge on how ionization dynamics of heavy atoms occur and how much they would affect neighboring atoms provides invaluable insight for successful x-ray molecular-imaging experiments.

In this article, we report a dramatic enhancement of ionization in molecules containing a mixture of light and heavy atomic species when they are exposed to ultraintense, ultrashort hard x-ray pulses.
The ionization of an iodobenzene (C$_6$H$_5$I) molecule irradiated by an XFEL pulse is theoretically investigated and compared with the ionization of an iodomethane (CH$_3$I) molecule and a xenon (Xe) atom, demonstrating that a larger molecule with a bigger electron reservoir gets more ionized and illustrating how induced charges are redistributed within the molecule.


In the calculation, we use the newly developed {\it ab initio} toolkit \textsc{xmolecule}~\cite{Hao15,Inhester16}.
This toolkit solves the electronic structure of molecules based on the Hartree-Fock-Slater method and simulates the multiphoton multiple ionization dynamics by solving a set of coupled rate equations.
The electronic structure, transition rates, and cross sections are calculated for every multiple-hole electronic configuration that may be formed during the interaction with an intense x-ray pulse.
The number of coupled rate equations involved in deep inner-shell ionization dynamics of C$_5$H$_6$I is about 9$\times$10$^{21}$, which can only be treated by employing the Monte Carlo on-the-fly technique~\cite{Fukuzawa13}.
For comparison, in the case of CH$_3$I, $\sim$2$\times$10$^{14}$ coupled rate equations had to be solved~\cite{Rudenko17}.
To achieve efficient calculation of molecular multiple-hole states, we use core-hole-adapted numerical atomic orbitals as basis functions, which are optimized for the respective atomic core-hole states by using \textsc{xatom}~\cite{Son11a}.
At every single ionization step, these basis functions and molecular orbitals are re-optimized.
This re-optimization models charge rearrangement within the molecule during the ionization dynamics.
The numerical basis set used in our calculations is constructed using a minimal basis set plus $5d$, $6s$, $6p$, and $6d$ basis functions on iodine.
For details of the re-optimization scheme and the basis set convergence, see the Appendices~\ref{appendix:MO} and \ref{appendix:basis_set}.
The x-ray photon energy is 8.3~keV, the pulse duration is 30~fs full-width at half-maximum (FWHM), and the fluence is varied up to 5$\times$10$^{12}$~photons/$\mu$m$^2$, corresponding to a peak intensity of 2$\times$10$^{19}$~W/cm$^2$.
In the ionization dynamics simulation, 13 fluence points are considered (0.1, 0.4, 0.8, 1.0, 1.2, 1.4, 2.0, 2.5, 3.0, 3.5, 4.0, 4.5 and 5.0$\times$10$^{12}$~photons/$\mu$m$^2$) and 39--112 Monte Carlo trajectories are obtained per fluence point.
The molecular geometry of iodobenzene is taken from Ref.~\cite{Brunvoll90} and that of iodomethane from Ref.~\cite{Johnson15}.
In the present work, nuclear dynamics are not considered because ultrafast ionization dynamics are not much affected by nuclear motion~\cite{Rudenko17}. 
See the Appendix~\ref{appendix:nuclear_dynamics} for further discussion of the effect of nuclear dynamics.
Resonance and relativistic effects are not included because their impact on the ionization dynamics at 8.3~keV is minimal~\cite{Rudek18}.


\begin{figure}[tbp]
\includegraphics[width=\figurewidth]{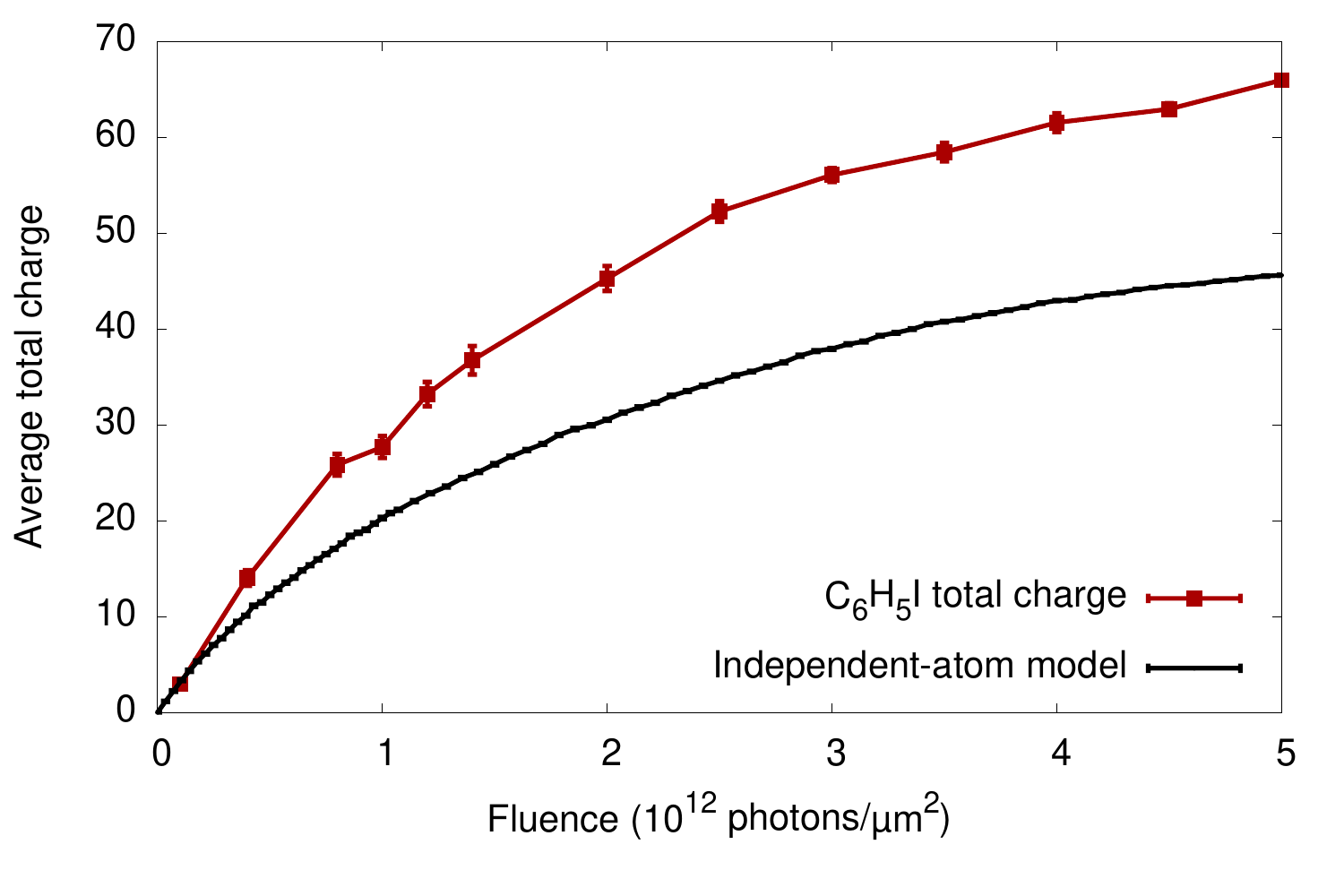}
\caption{(Color online) Average total molecular charge of iodobenzene calculated using the full molecular model and within the independent-atom model. 
The error bars represent the statistical uncertainty of the data.}
\label{fig:total_charge}
\end{figure}

In Fig.~\ref{fig:total_charge}, we show the average total molecular charge after the interaction of iodobenzene with the XFEL pulse as a function of fluence, calculated for two different models. 
The iodobenzene total charge (red) indicates the full molecular model given by \textsc{xmolecule}, while the independent-atom model (black) shows the sum of the individual atomic charges of six carbons, five hydrogens, and one iodine, each of which is calculated using \textsc{xatom}.
In the low-fluence regime (around 10$^{11}$~photons/$\mu$m$^2$), there is almost no difference between the full molecular model and the independent-atom model.
As the fluence increases, however, the discrepancy between them becomes noticeable and the molecular effect clearly enhances the total molecular charge.
At the maximum fluence we considered (5$\times$10$^{12}$~photons/$\mu$m$^2$), the average total molecular charge is +66, whereas the independent-atom model gives +46, so the discrepancy between them is 20 charges.
This dramatic enhancement of ionization can be explained by charge-rearrangement-enhanced x-ray ionization of molecules (CREXIM)~\cite{Rudenko17}.
Hard x-ray photons are absorbed almost exclusively by the iodine atom, because its cross section is much larger than that of the light atoms ($\sigma_\text{I}$=0.052~Mb, $\sigma_\text{C}$=7.5$\times$10$^{-5}$~Mb, $\sigma_\text{H}$=8.5$\times$10$^{-9}$~Mb at 8.3~keV).
A deep inner-shell vacancy after photoionization at iodine induces an Auger cascade, ejecting several electrons and yielding a highly charged iodine ion.
The resulting charge imbalance between iodine and neighboring light atoms drives charge rearrangement via the chemical bonding network, i.e., electron transfer from the phenyl group to the iodine site.
Those electrons, which would not be directly photoionized at the light-atom sites, are available for further photoionization at the iodine site.
Repeated photoionization at iodine and intramolecular charge transfer from the light atoms to iodine result in more ionization than in the case of independent atoms.
The CREXIM mechanism has been proposed for H$_2$O~\cite{Inhester16} and quantitatively confirmed with experiment for CH$_3$I~\cite{Rudenko17}. 
The CREXIM effect manifests itself not only in the produced charges, but also in electron and fluorescence spectra~\cite{Schaefer18}.
In our calculations, charge rearrangement is described via instantaneous orbital relaxation upon sequential ionization and molecular Auger decay~\cite{Inhester16}. 
We do not take into account coherent electron motion~\cite{Kuleff05,Remacle06}. 
It must be expected that tracing over the large number of electrons ejected suppresses the fingerprints of electronic coherence.


\begin{figure}[tbp]
\includegraphics[width=\figurewidth]{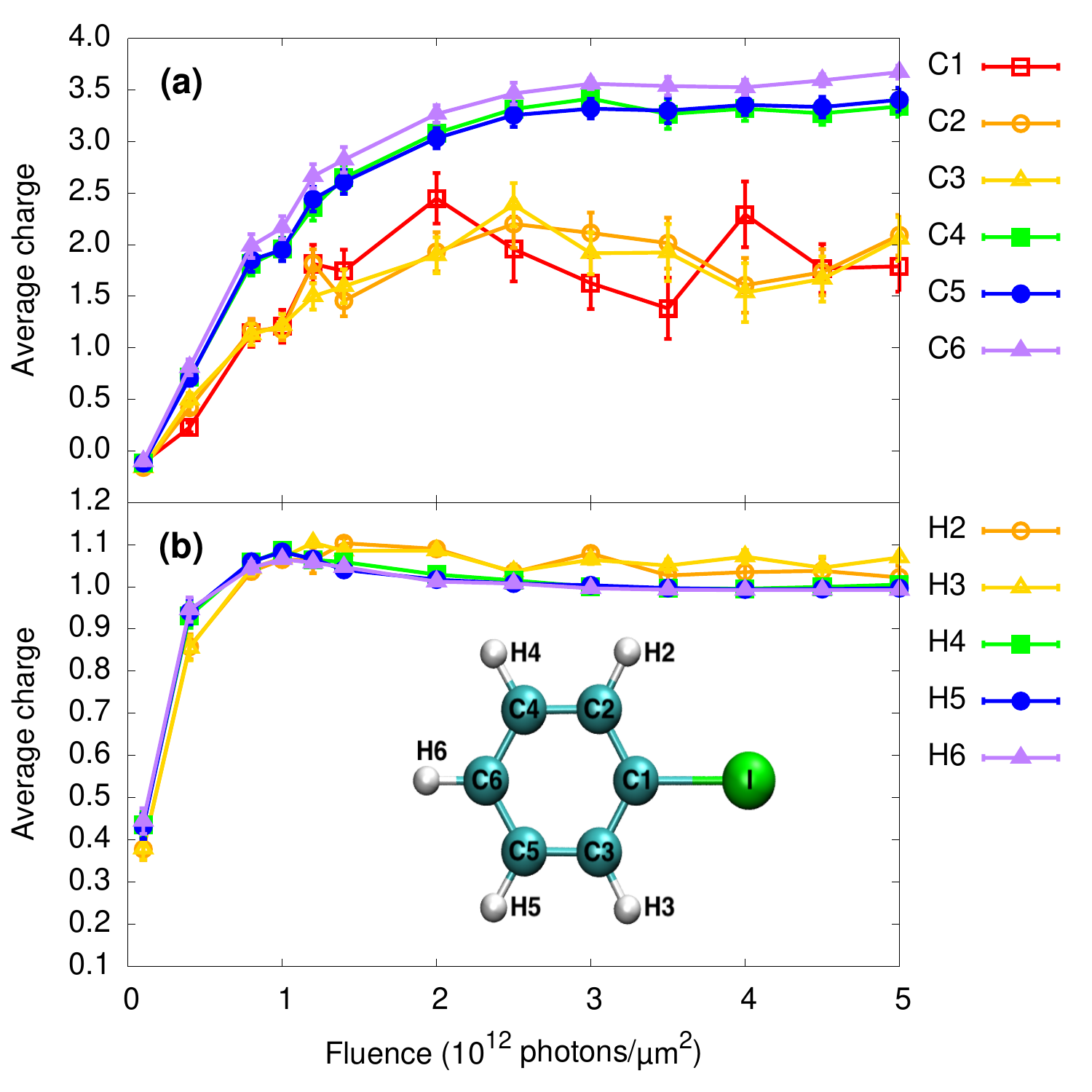}
\caption{(Color online) Average partial charge of (a) carbon and (b) hydrogen as a function of fluence for C$_6$H$_5$I molecules. 
The error bars represent the statistical uncertainty of the data.
The atomic labels employed are indicated in the shown structure of C$_6$H$_5$I.}
\label{fig:phenyl_group}
\end{figure}

To further inspect the CREXIM mechanism, we examine how the positive charges (holes) are redistributed inside the benzene ring after electrons have been transferred to the iodine atom.
Figure~\ref{fig:phenyl_group} shows the partial charges of (a) carbons and (b) hydrogens as a function of fluence, in order to identify the charge distribution in the phenyl group.
Note that all these positive charges on the light atoms are almost exclusively produced by charge rearrangement rather than by direct photoionization.
C1 indicates the carbon atom attached to iodine.
C2 and C3 are the next closest carbon atoms to iodine.
C4 and C5 are farther away, and C6 represents the carbon atom farthest from iodine.
Figure~\ref{fig:phenyl_group}(a) shows that the partial charges divide the carbon atoms into two groups: those closer to iodine (C1, C2, and C3) and those farther away from iodine (C4, C5, and C6).
Surprisingly, the carbons at the remote end get more ionized than the carbons at the near end.
This is somewhat counterintuitive because one might expect that charge transfer would occur more effectively when the carbon atom involved in the charge transfer is closer to iodine.
However, high positive charges on carbon and iodine have a strong Coulomb repulsion, and an energetically more stable configuration is obtained when they are separated by a larger internuclear distance.
Therefore, after charge redistribution inside the benzene ring, more positive partial charges are formed at carbons farther away from iodine.
The charge fluctuation of the carbon group at the near end might be attributed to low statistics reflected by large error bars.
In Fig.~\ref{fig:phenyl_group}(b), there is no difference in hydrogen partial charges for different locations.
As the fluence increases, the hydrogen atoms always get charged to +1, no matter which carbon they are attached to.


\begin{figure}[tbp]
\includegraphics[width=\figurewidth]{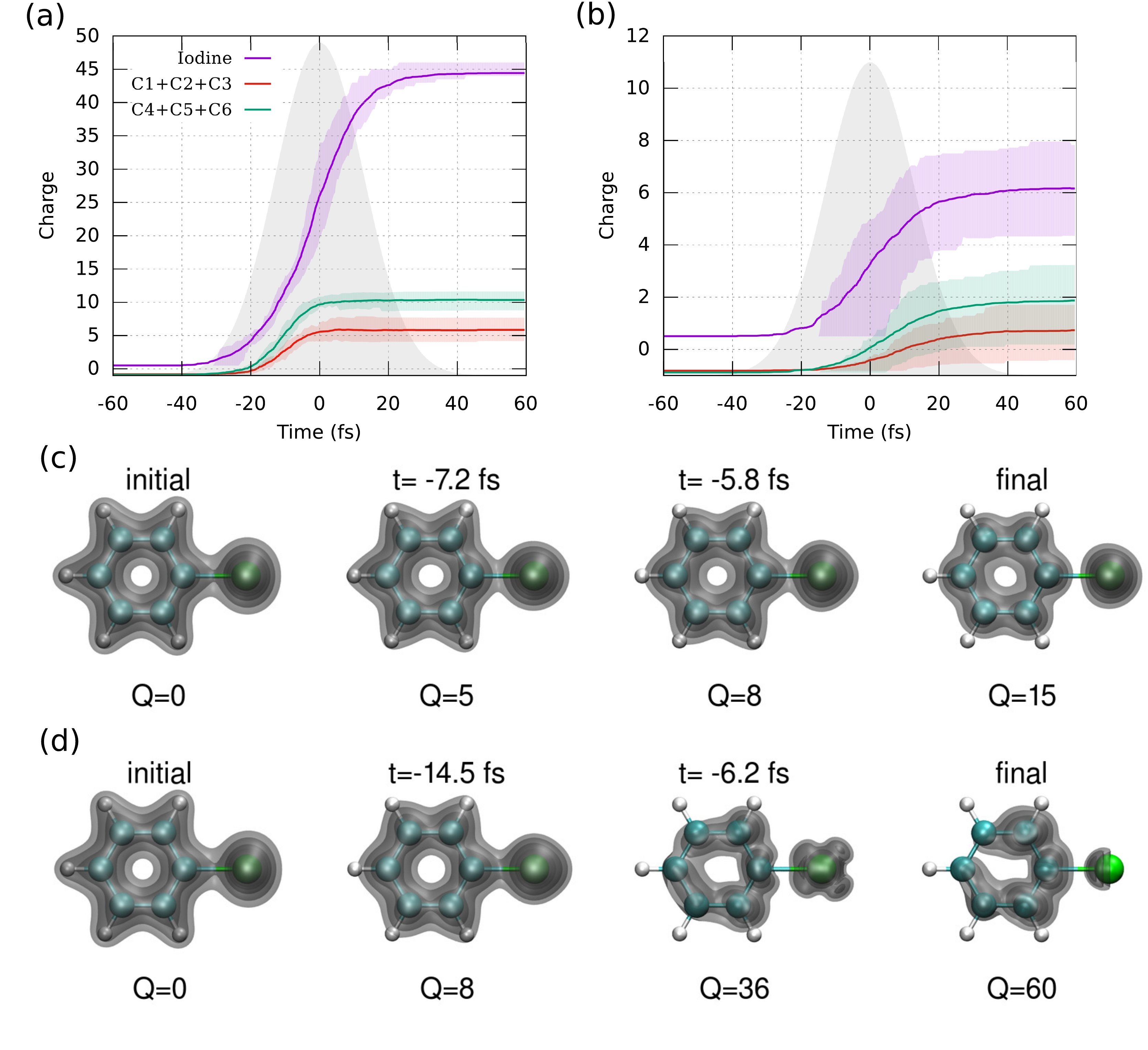}
\caption{(Color online) Time evolution of partial charges and snapshots of electron density of iodobenzene at selected times during the pulse exposure.
(a,b) Partial charges of iodine (magenta) and summed partial charges of the carbon atoms, C1+C2+C3 (red) and C4+C5+C6 (green), as a function of time at the fluence of (a) 5$\times$10$^{12}$ photons/$\mu$m$^2$ and (b) 4$\times$10$^{11}$ photons/$\mu$m$^2$.
The solid lines show the average partial charge, the colored areas indicate the range between the upper and lower quartiles of the partial charge distribution. 
The gray area indicates the envelope of the pulse.
(c,d) Snapshots of the electron density at selected times $t$ for typical trajectories at the fluence of (c) 4$\times$10$^{11}$ photons/$\mu$m$^2$ and (d) 5$\times$10$^{12}$ photons/$\mu$m$^2$. 
$Q$ indicates the total charge of the molecule at time $t$.
}
\label{fig:time_evolution_and_snapshots}
\end{figure}

In the upper panels of Fig.~\ref{fig:time_evolution_and_snapshots}, we show the time evolution of the partial charges of iodine and the carbons as iodobenzene is irradiated by hard x-ray pulses at (a) high fluence (5$\times$10$^{12}$ photons/$\mu$m$^2$) and (b) low fluence (4$\times$10$^{11}$ photons/$\mu$m$^2$).
The carbon partial charges are summed into two groups, the one closer to iodine (C1+C2+C3; red) and the other farther from iodine (C4+C5+C6; green).
The Gaussian-shaped pulse is also plotted with a gray area as a reference. 
During the whole interaction process of the iodobenzene molecule with the x-ray pulse, the iodine charge increases continuously and arrives at the maximum towards the end of the pulse. 
After the end of the pulse the charge stays constant.
In Fig.~\ref{fig:time_evolution_and_snapshots}(a), the carbon atoms reach their final charges much earlier than the iodine charge and well before the peak of the x-ray pulse, which is a similar trend as observed for iodomethane under the same x-ray beam conditions (Fig.~4 in Ref.~\cite{Rudenko17}).
For the low-fluence case in Fig.~\ref{fig:time_evolution_and_snapshots}(b), the time scales of iodine and carbon charge dynamics are similar, but their final charges are not as high as in the high-fluence case, which will result in slower Coulomb explosion dynamics.
Therefore, for the short pulses considered here, the charge rearrangement dynamics are completed before considerable molecular dissociation occurs (see the Appendix~\ref{appendix:nuclear_dynamics}), which makes CREXIM relatively insensitive to nuclear dynamics.

The lower panels of Fig.~\ref{fig:time_evolution_and_snapshots} show snapshots of the electron density for the (c) low- and (d) high-fluence cases.
One can clearly see that all hydrogens lose one electron for both cases and the carbons farther away from iodine lose more electrons than the ones closer to iodine for the high-fluence case.
The evolution of the electron density illustrates that, for a medium-sized molecule containing a heavy atom like iodobenzene, the electronic radiation damage occurs globally, i.e., over the whole molecule, rather than locally at the heavy atom and its vicinity.
Note that the final states of (c) and (d) are far from the electronic ground state for the corresponding charge state.
For instance, the electronic ground state of $Q$=60 [the final charge of (d)] is a state in which all remaining electrons are located at the iodine site, due to the strong Coulomb potential from iodine.
The electron density plot of the final state of (d), however, shows a significant contribution of electron density on the phenyl group.
Since there are huge potential barriers among highly charged ions, electrons from the phenyl group cannot be fully transferred to iodine.
Also note that the electron density snapshots shown in Figs.~\ref{fig:time_evolution_and_snapshots}(c) and (d) are obtained from one exemplary trajectory each.
The electron density has broad fluctuations for different trajectories, as may be anticipated based on the fluctuations of the partial charges in Figs.~\ref{fig:time_evolution_and_snapshots}(a) and (b).


\begin{figure}[tbp]
\includegraphics[width=\figurewidth]{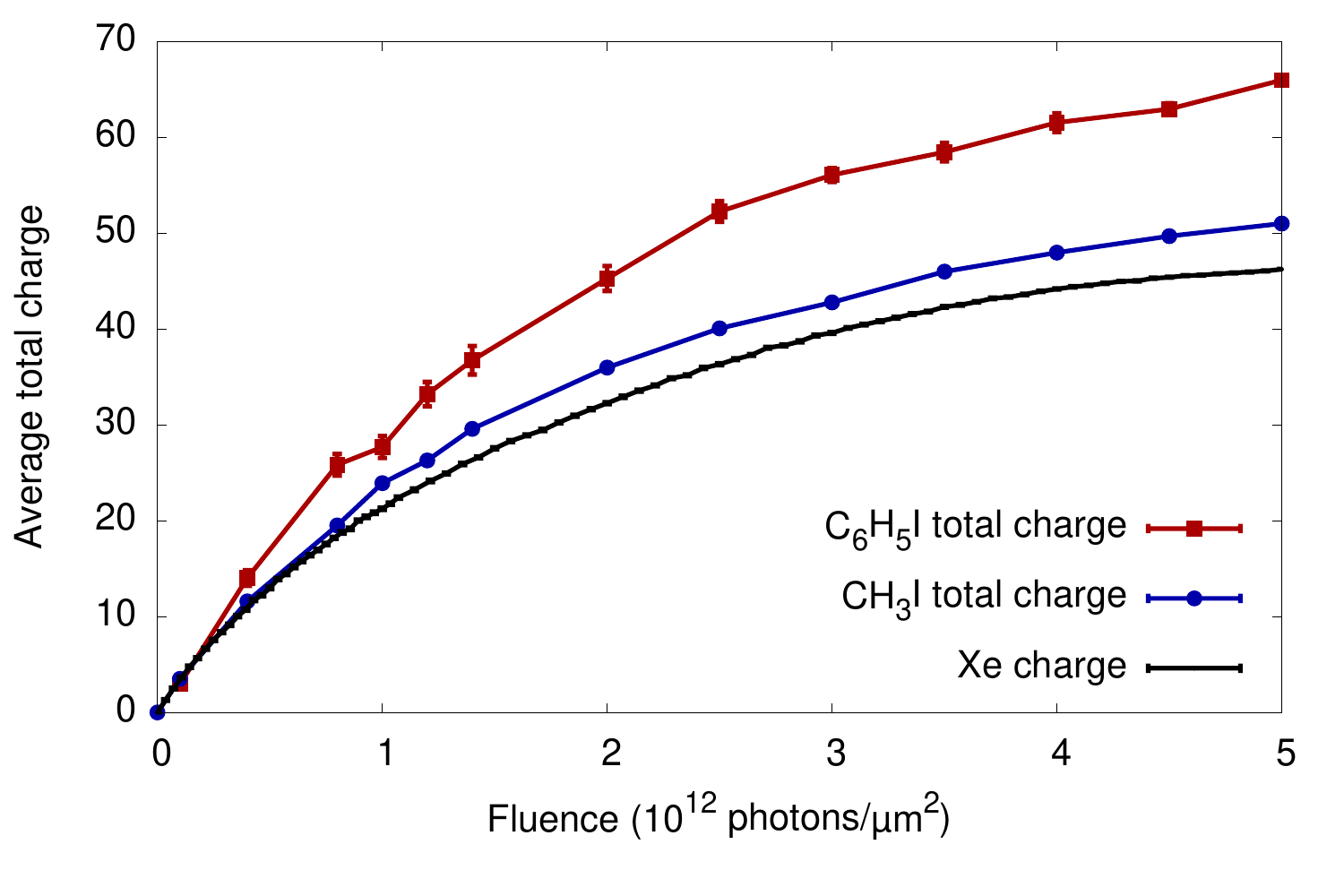}
\caption{(Color online) Average final charge of xenon, iodomethane, and iodobenzene at 8.3~keV as a function of fluence.}\label{fig:comparison_various_systems}
\end{figure}

It has been speculated that the CREXIM effect becomes stronger for larger polyatomic systems where more electrons can be transferred from neighboring atoms to the x-ray-absorbing site~\cite{Rudenko17}.
Bearing in mind that absorption of hard x rays by light atoms is almost negligible, the calculated absorption cross sections for Xe, CH$_3$I, and C$_6$H$_5$I at 8.3~keV are very similar (0.056~Mb, 0.052~Mb, and 0.053~Mb, respectively).
The experimental charge-state distributions (CSD) of the xenon ions and the iodine ions emitted from CH$_3$I and C$_6$H$_5$I under the same experimental conditions were very similar (Fig.~1 in Ref.~\cite{Rudenko17}).
Combined with the fact that very high iodine charges are detected along with carbon charges of +4 for the CH$_3$I case, the similarity of the CSD of heavy-atom ion fragments implies that the sum of all fragment charges for C$_6$H$_5$I could be much larger than that for CH$_3$I, but no verification had been made until now.
The present study confirms this earlier speculation by demonstrating a stronger ionization for the larger molecule.
Figure~\ref{fig:comparison_various_systems} shows the average total molecular charge as a function of fluence for Xe, CH$_3$I, and C$_6$H$_5$I.
The average charges for the three species are similar in the low-fluence regime ($<$5$\times$10$^{11}$~photons/$\mu$m$^2$), while they deviate from each other as the fluence increases.
At the highest fluence used in our calculation (5$\times$10$^{12}$ photons/$\mu$m$^2$), the average charge of Xe is +46, while the average total molecular charge of CH$_3$I reaches +54 (17\% enhancement in comparison with the Xe case)~\cite{Rudenko17}.
Moreover, the average total molecular charge of C$_6$H$_5$I is +66 (43\% enhancement).
This enormous ionization enhancement is due to the larger number of electrons on the phenyl (--C$_6$H$_5$) group in comparison to the methyl (--CH$_3$) group.
Our results demonstrate that the larger the molecule, the stronger is the CREXIM effect and the more severe is the high-intensity radiation damage to the molecule.

The CREXIM effect is a new aspect for a quantitative understanding of electronic radiation damage to molecules in x-ray imaging, which is fundamentally different from electronic radiation damage to isolated atoms.
Another important source of radiation damage in extended systems is electron impact ionization~\cite{Vinko12,Murphy14,Ziaja15}.
X-ray multiphoton absorption by a heavy atom within a molecule produces multiple photo- and Auger electrons with kinetic energies ranging from a few hundred eV to a few keV, which cause impact ionization to neighboring atoms.
These two effects of impact ionization and CREXIM enhance the degree of ionization in extended systems.
Here, the heavy atom functions both as an ionization catalyst through CREXIM and as an electron source for impact ionization.
It is expected that both CREXIM and impact ionization effects increase for larger molecules.
The former is demonstrated by the present work and the latter is because its probability per electron is given by $p = 1 - e^{-x / \lambda}$, where $x$ is the distance traveled and $\lambda$ is the mean free path.
For neutral bulk protein, the effective electron mean free path is about 10--20~\AA\ for electron kinetic energies in the range from 100 to 1000~eV~\cite{Ziaja15}, and the impact ionization probability is less than $\sim$40\% when the distance is less than half of the mean free path.
In our study, the size of iodobenzene is about 6~\AA, which is smaller than the effective mean free path; thus no impact ionization is considered.
Hence, it would seem that CREXIM dominates on short length scales, while impact ionization dominates on longer length scales.
Thus, one might expect a transition of the radiation damage mechanism from CREXIM to impact ionization, as a function of the distance from the heavy atom position.
The present work, however, indicates that CREXIM does not show an inverse relation with respect to the distance.
Instead, we find the opposite trend: the longer the distance from the heavy atom, the higher the positive charge of the carbons in the benzene ring.
In order to better understand radiation damage dynamics of larger heavy-atom-containing molecules, it will be imperative to include impact ionization and to investigate the distance-dependence and the chemical-environment-dependence of the CREXIM effect in a systematic manner.


In summary, we have performed first-principles calculations of ionization dynamics of iodobenzene irradiated by an ultrashort and ultraintense hard x-ray pulse.
We have found a significant enhancement of molecular ionization when a molecule consists of a mixture of light and heavy atoms.
The mechanism of this ionization enhancement is called CREXIM, which is characterized by sequential multiphoton ionization at a heavy atom site and intramolecular electron transfer from light atoms to the heavy atom.
After electron transfer, the positive charges formed in the benzene ring are redistributed such that the carbons positioned at a longer distance from iodine are more charged than those located closer to iodine.
Through comparison of the average total charges of Xe, CH$_3$I, and C$_6$H$_5$I at high x-ray intensity, we have found that the CREXIM effect becomes stronger as the molecular size increases.
This CREXIM effect has not yet been taken into account in existing computer programs for simulating the interaction of x rays with matter~\cite{Peltz12,Hau-Riege13a,Jurek16,Ho17}.
Our work presents illuminating insights as to how electronic radiation damage occurs in heavy-atom-containing molecules, and therefore, we expect it to be of significance to XFEL applications.

\begin{acknowledgments}
This work has been supported by the NSF of China under Grants No.~117.
%
\end{acknowledgments}

\appendix
\setcounter{figure}{0}
\renewcommand\thefigure{A\arabic{figure}}
\section{Re-optimization of molecular orbitals at each ionization step}\label{appendix:MO}

During the consecutive ionization steps, molecular orbitals (MOs) are re-optimized for the new electronic configuration employing a variant of the maximum overlap method~\cite{Hao15}.
For the repeated self-consistent field (SCF) calculation, we employ averaged fractional occupation numbers for nearly degenerate MOs, which are particularly useful for deep inner-shell vacancies on the iodine atom. 
In the case that the SCF calculation fails to converge, we try to smoothly adapt the occupation number to the new configuration in fractional steps.
If this procedure still fails, we additionally employ averaged fractional occupation numbers for open inner shells that have become localized on the iodine atom during multiple ionization. 
If all these procedures fail to obtain a converged solution in the SCF procedure, then we proceed with MOs optimized from the previous step.
The last case happens in about 15\% of all the ionization steps in our calculations.

\section{Basis set convergence}\label{appendix:basis_set}
\begin{figure}[h]
\centering
\includegraphics[width=0.5\figurewidth]{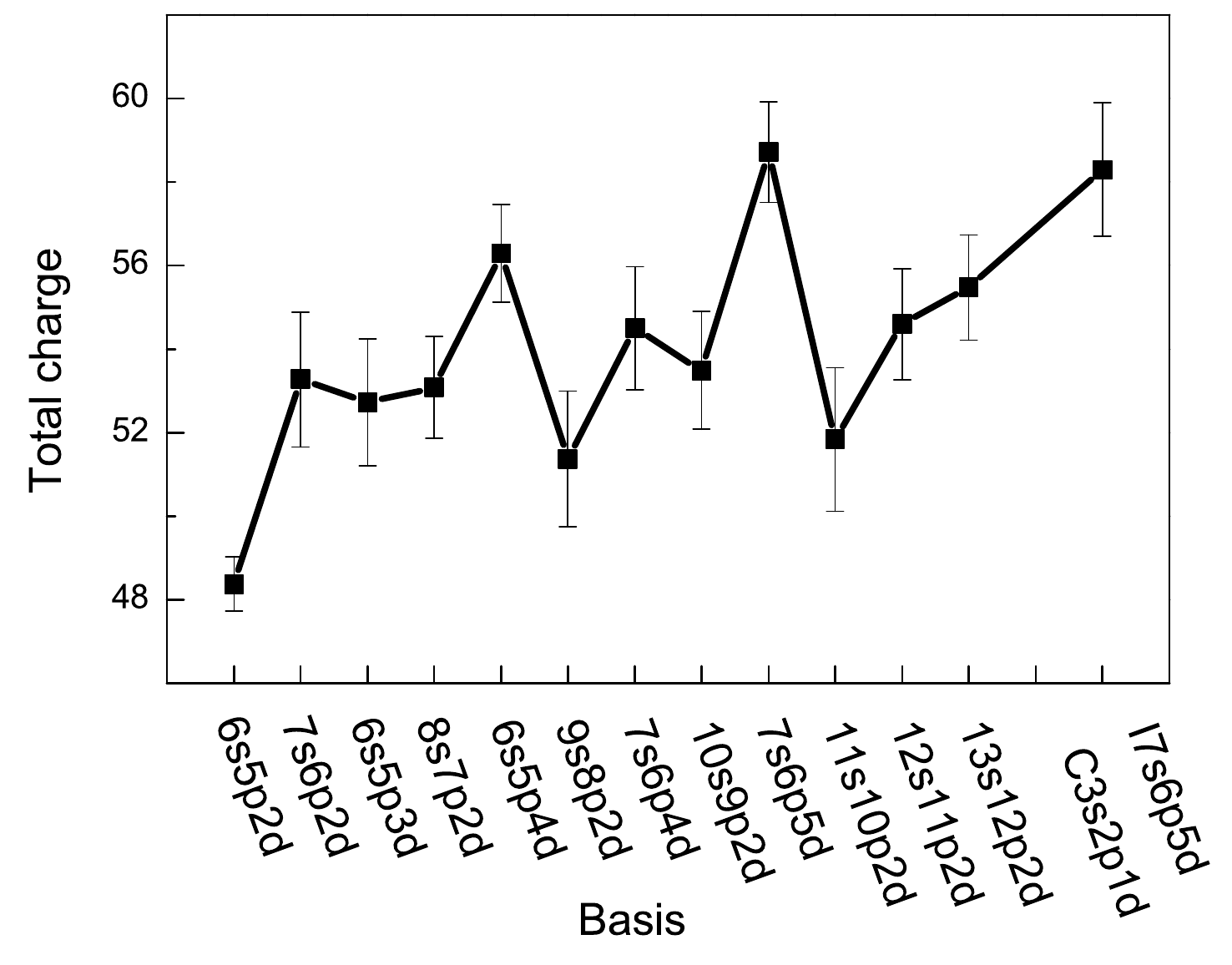}
\caption{Basis-set-dependence of the total molecular charge of C$_6$H$_5$I calculated at 3$\times$10$^{12}$~photons/$\mu$m$^2$. 
The basis-set label of AsBpCd indicates that the numbers of $s$, $p$, and $d$ atomic subshells are A, B, and C, respectively.
The basis sets labeled on the $x$-axis are used for iodine, while the minimal basis set is used for carbon and hydrogen, except for the last one where 7s6p5d for iodine and 3s2p1d for carbon are used.
}\label{fig:basis_set}
\end{figure}

The CREXIM effect is driven by intramolecular electron transfer via the chemical bonding network.
During multiple ionization dynamics induced by an intense x-ray pulse, highly charged atomic ions are formed, and the atomic orbitals, which we employ as basis functions, shrink in size.
Hence, potentially small overlaps among atomic orbitals from highly charged atomic species require using sufficiently diffuse basis functions.
In Fig.~\ref{fig:basis_set}, we test the convergence of the molecular total charge calculated at 3$\times$10$^{12}$~photons/$\mu$m$^2$ with different basis sets.
We start with the basis set 6s5p2d, which is constructed with the minimal basis set for iodine (5s4p2d) plus atomic orbitals corresponding to the $6s$ and $6p$ subshells.
The number of basis functions is calculated as follows: one function per $s$-type atomic subshell, three functions per $p$-type atomic subshell, and five functions per $d$-type atomic subshell.
For example, the iodine label of 6s5p2d gives $N_\text{I}$=31 ($=6 + 5\times3 + 2\times5$) basis functions for iodine.
Along the $x$-axis of Fig.~\ref{fig:basis_set}, the number of basis functions used for iodine increases according to the iodine basis-set label.
A general trend is that the total charge increases as the number of basis functions increases.
However, the calculated total charge peaks with 6s5p4d and 7s6p5d, illustrating that it is sensitive to the inclusion of $d$-type atomic orbitals, i.e., polarization functions for the I--C bonding.
Adding more diffuse $s$-type and $p$-type functions, for example, 10s9p2d, 11s10p2d, 12s11p2d, and 13s12p2d, does not alter the calculated total charge that much.
It is worthwhile to note that the calculated total charge is about +54 with a statistical error of $\pm$2, and it fluctuates by about $\pm$6 ($\sim$11\%) depending on the basis set.
Therefore, we chose 6s5p4d for iodine, corresponding to the minimal basis set (5s4p2d) plus one more $s$-type ($6s$), one more $p$-type ($6p$), and two more $d$-type ($5d$ and $6d$) subshells, in all the calculations shown in the main text.
The minimal basis set is used for carbon (2s1p) and hydrogen (1s), so the total number of basis functions used in all the calculations for C$_6$H$_5$I is 76 ($=6 N_\text{C} + 5 N_\text{H} + N_\text{I} = 6\times5 + 5\times1 + 41$).

\section{Effect of nuclear dynamics}\label{appendix:nuclear_dynamics}

For a large system like C$_6$H$_5$I, carrying out simulations including nuclear dynamics is beyond the current computational capabilities of \emph{ab initio} calculations.
For the smaller CH$_3$I molecule, however, it is possible to conduct the full simulation including the nuclear dynamics and to directly compare the results with and without nuclear dynamics, as demonstrated in the Extended Data Figure~3 in Ref.~\cite{Rudenko17}.
As can be seen in that figure, the effect of nuclear dynamics on the total molecular charge is marginal, which is attributed to the fact that the charge rearrangement occurs at very early times during the pulse, where the distance between C and I is still relatively small.

\begin{figure}[h]
\centering
\includegraphics[width=0.5\figurewidth]{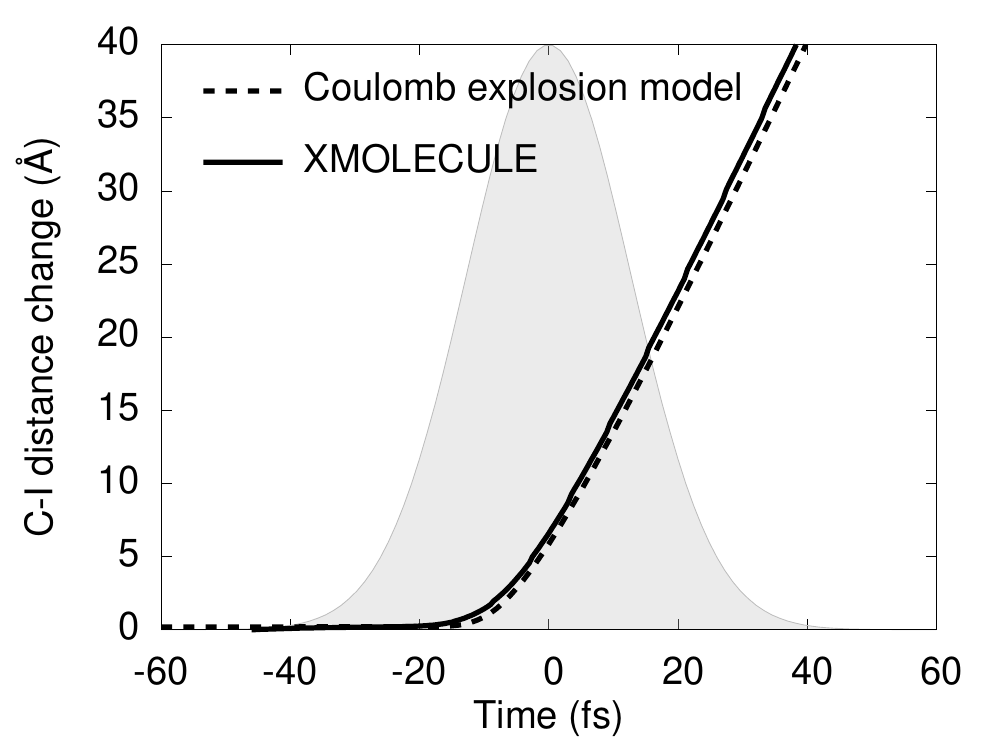}
\caption{Time evolution of the average C--I bond distance.
The solid line is obtained from the full ionization and nuclear dynamics calculation using \textsc{xmolecule}~\cite{Rudenko17}, while the dashed line is computed from the simplified Coulomb explosion model. 
The fluence is fixed at $5\times10^{12}$~photons/$\mu$m$^2$.
The shaded gray area represents a Gaussian pulse envelope of 30~fs FWHM.
\label{fig:CH3I_nuc_dynamics}}
\end{figure}

In order to inspect the interplay between nuclear dynamics and ionization dynamics for larger systems, we here employ a simplified Coulomb explosion model, in which the atomic positions are propagated classically with Coulomb forces among the charges created on the atoms.
Instead of employing an instantaneous build-up of atomic charges, which would overestimate the speed of the dissociation dynamics, we employ the time-dependent partial charges on the atoms extracted from \textsc{xmolecule} calculations and plug them into the Coulomb explosion model as an input.
To validate this approach, we compare its results with the full ionization and nuclear dynamics calculation for CH$_3$I~\cite{Rudenko17}.
Figure~\ref{fig:CH3I_nuc_dynamics} shows the time-dependent C--I bond distance change at $5\times10^{12}$~photons/$\mu$m$^2$ for both the full simulation (\textsc{xmolecule}) and the Coulomb explosion model.
The full dynamics calculation is taken from Fig.~4(b) in Ref.~\cite{Rudenko17}, and the Coulomb explosion model is based on the gradual build-up of average partial charges extracted from Fig.~4(a) in Ref.~\cite{Rudenko17}.
We ignore negative partial charges that occur in the initially neutral molecule.
As shown in Fig.~\ref{fig:CH3I_nuc_dynamics}, the simplified Coulomb explosion model reproduces the full dynamics calculation in the CH$_3$I case.

\begin{figure}[h]
\centering
\includegraphics[width=\figurewidth]{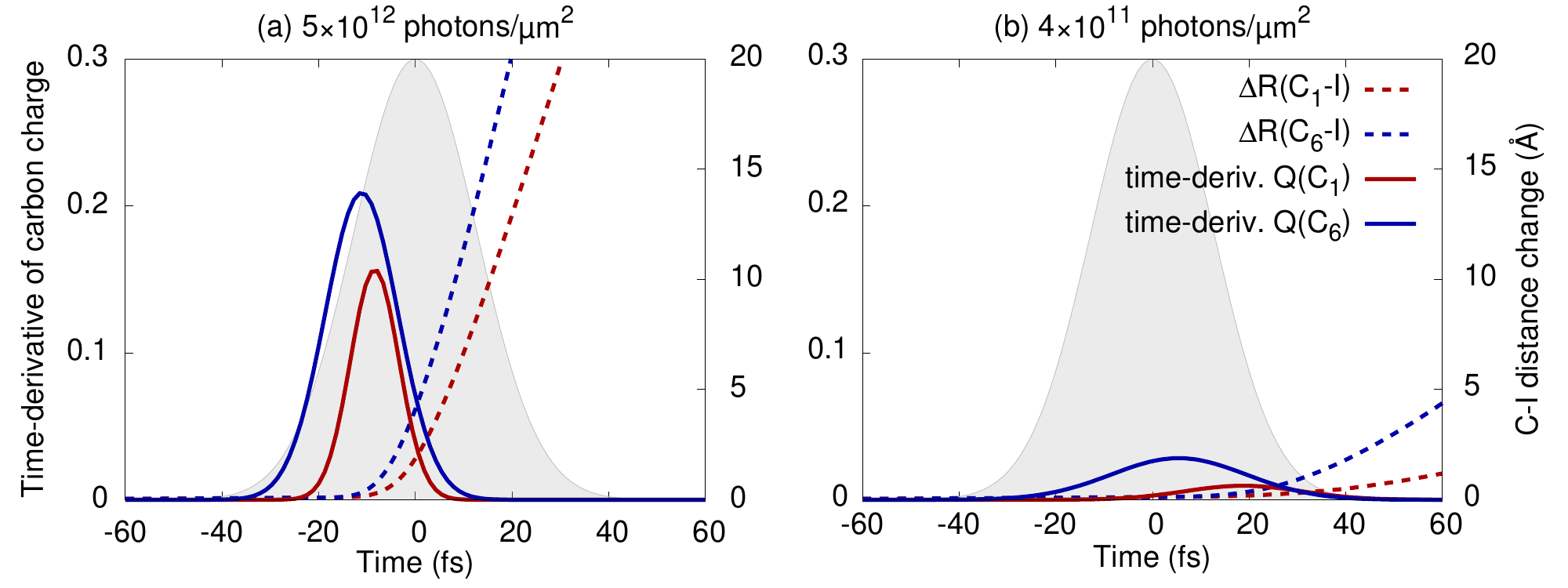}
\caption{Time evolution of charge rearrangement and estimated nuclear motion of C$_6$H$_5$I.
The former is represented by the time derivative of the average charge of carbon (solid line) and the latter is calculated using the Coulomb explosion model (dashed line).
The red color is for C$_1$ and the blue color is for C$_6$.
The fluence used is (a) $5\times10^{12}$~photons/$\mu$m$^2$ and (b) $4\times10^{11}$~photons/$\mu$m$^2$.
\label{fig:C6H5I_nuc_dynamics}
}
\end{figure}

Based on the good agreement for CH$_3$I, we perform the same Coulomb explosion model calculation for C$_6$H$_5$I, where the full dynamics calculation is not available.
Figure~\ref{fig:C6H5I_nuc_dynamics} displays charge rearrangement dynamics and estimated nuclear dynamics of C$_6$H$_5$I together.
Here we focus on the C--I bond distance of the carbon atom close to the iodine atom, C$_1$ (red), and the carbon atom at the opposite end of the molecule, C$_6$ (blue).
The time-dependent C--I bond distances, as plotted using dashed lines in Figs.~\ref{fig:C6H5I_nuc_dynamics}(a) $5\times10^{12}$~photons/$\mu$m$^2$ and (b) $4\times10^{11}$~photons/$\mu$m$^2$, are calculated using the Coulomb explosion model with the charging-up information of all individual atoms---the iodine charge and the sum of carbon charges are shown in Figs.~3(a) and (b).
Note that at $5\times10^{12}$~photons/$\mu$m$^2$ the distance of C$_1$--I in the C$_6$H$_5$I molecule evolves considerably slower than the corresponding C--I distance in CH$_3$I.
This is because the charge is effectively distributed within the benzene ring and the individual atom feels a much lower Coulomb force.
Furthermore, the C$_1$ atom is partially pushed back by the charge that is built up on the other carbon atoms.

\begin{figure}[h]
\centering
\includegraphics[width=\figurewidth]{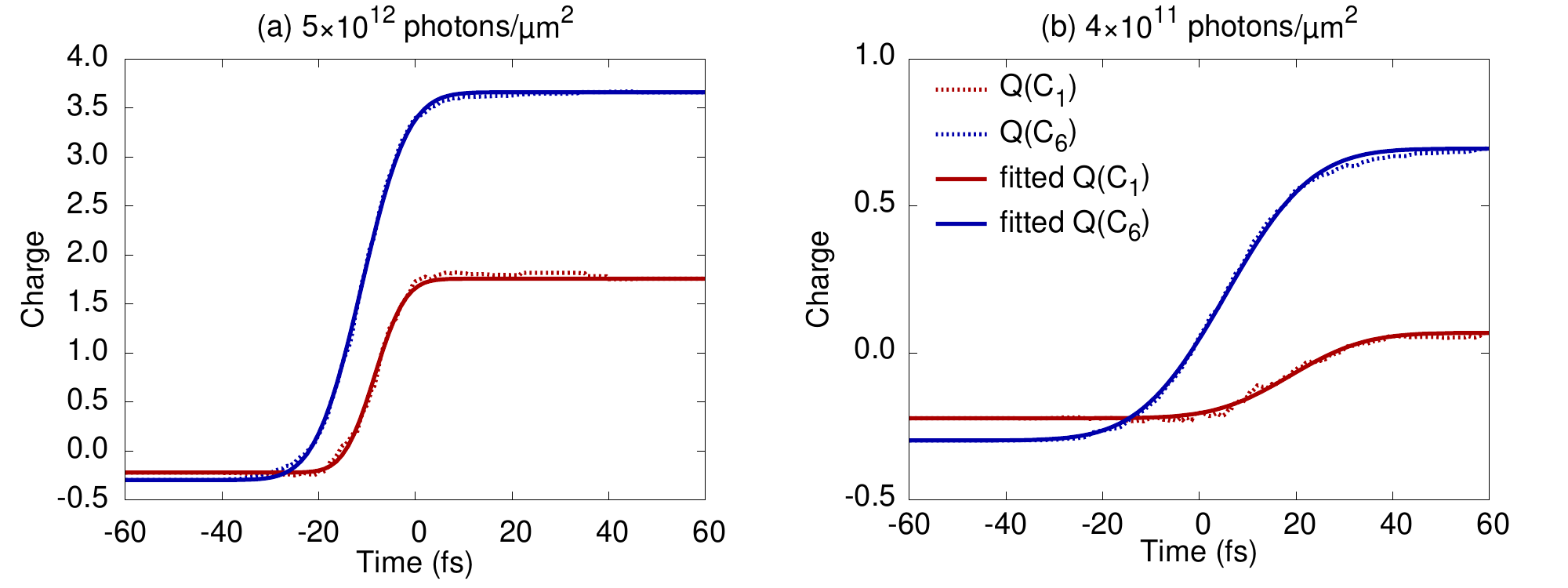}
\caption{Time evolution of average C$_1$ and C$_6$ partial charges.
The dotted line is extracted from \textsc{xmolecule} calculations, and the solid line represents the fitting function from Eq.~(\ref{eq:fitting}).
\label{fig:C6H5I_fitting}
}
\end{figure}

In the same figure, the charge rearrangement dynamics are represented by the time derivative of the partial carbon charges.
To achieve better visibility, the time evolution of the partial charges of the C$_1$ and C$_6$ components from Figs.~3(a) and (b) are fitted to the following function,
\begin{equation}\label{eq:fitting}
Q_{t_0,\tau}(t) = q_i + \frac{q_f - q_i}{2} \left[ 1 + \text{erf}\left( \frac{ \sqrt{4 \ln 2} (t-t_0) }{\tau} \right) \right],
\end{equation}
where $q_i$ is the initial charge as $t \rightarrow -\infty$ and $q_f$ is the final charge as $t \rightarrow +\infty$.
Here, the time shift $t_0$ and the width $\tau$ (FWHM) are varied to fit the ionization dynamics of C$_1$ and C$_6$.
Figure~\ref{fig:C6H5I_fitting} shows how the fitting procedure works for the C$_1$ and C$_6$ partial charges.
Note that a similar ionization-dynamics modeling approach was proposed in Ref.~\cite{Motomura15} with a different fitting function.
The time derivative of Eq.~(\ref{eq:fitting}) is given by
\begin{equation}
\frac{d}{d t} Q_{t_0,\tau}(t) = \frac{q_f - q_i}{\tau} \sqrt{ \frac{4 \ln 2}{\pi} } e^{ - 4 \ln 2 (t-t_0)^2 / \tau^2 },
\end{equation}
which is plotted using solid lines in Figs.~\ref{fig:C6H5I_nuc_dynamics}(a) and (b).
From this time-derivative plot, one can easily read out when the charging-up dynamics of carbon starts and saturates, i.e., the time scale of the charge rearrangement dynamics.
For example, the charge of C$_6$ in Fig.~\ref{fig:C6H5I_nuc_dynamics}(a) is developed starting at $\sim$$-30$~fs and ending at $\sim$$10$~fs.
Thus, the charge transfer to C$_6$ at $5\times10^{12}$~photons/$\mu$m$^2$ is terminated at $\sim$$10$~fs, when the C--I distance is elongated by $\sim$$10$~\AA.
For C$_1$, the charge transfer is terminated earlier and the nuclear dynamics is slower than for C$_6$.
For both cases, the charge rearrangement occurs at very early times before the molecule breaks apart considerably.
At low fluence of $4\times10^{11}$~photons/$\mu$m$^2$, the charge rearrangement occurs late in the pulse, but the bond distance is increased by only a few \AA.
Therefore, for the x-ray beam parameters under consideration, it is unlikely that the nuclear motion influences the charge rearrangement dynamics, and thus the CREXIM effect is unlikely to be affected by the nuclear dynamics.

%
%

\begin{thebibliography}{42}%
\makeatletter
\providecommand \@ifxundefined [1]{%
 \@ifx{#1\undefined}
}%
\providecommand \@ifnum [1]{%
 \ifnum #1\expandafter \@firstoftwo
 \else \expandafter \@secondoftwo
 \fi
}%
\providecommand \@ifx [1]{%
 \ifx #1\expandafter \@firstoftwo
 \else \expandafter \@secondoftwo
 \fi
}%
\providecommand \natexlab [1]{#1}%
\providecommand \enquote  [1]{``#1''}%
\providecommand \bibnamefont  [1]{#1}%
\providecommand \bibfnamefont [1]{#1}%
\providecommand \citenamefont [1]{#1}%
\providecommand \href@noop [0]{\@secondoftwo}%
\providecommand \href [0]{\begingroup \@sanitize@url \@href}%
\providecommand \@href[1]{\@@startlink{#1}\@@href}%
\providecommand \@@href[1]{\endgroup#1\@@endlink}%
\providecommand \@sanitize@url [0]{\catcode `\\12\catcode `\$12\catcode
  `\&12\catcode `\#12\catcode `\^12\catcode `\_12\catcode `\%12\relax}%
\providecommand \@@startlink[1]{}%
\providecommand \@@endlink[0]{}%
\providecommand \url  [0]{\begingroup\@sanitize@url \@url }%
\providecommand \@url [1]{\endgroup\@href {#1}{\urlprefix }}%
\providecommand \urlprefix  [0]{URL }%
\providecommand \Eprint [0]{\href }%
\providecommand \doibase [0]{http://dx.doi.org/}%
\providecommand \selectlanguage [0]{\@gobble}%
\providecommand \bibinfo  [0]{\@secondoftwo}%
\providecommand \bibfield  [0]{\@secondoftwo}%
\providecommand \translation [1]{[#1]}%
\providecommand \BibitemOpen [0]{}%
\providecommand \bibitemStop [0]{}%
\providecommand \bibitemNoStop [0]{.\EOS\space}%
\providecommand \EOS [0]{\spacefactor3000\relax}%
\providecommand \BibitemShut  [1]{\csname bibitem#1\endcsname}%
\let\auto@bib@innerbib\@empty
\bibitem [{\citenamefont {Marangos}(2011)}]{Marangos11}%
  \BibitemOpen
  \bibfield  {author} {\bibinfo {author} {\bibfnamefont {J.~P.}\ \bibnamefont
  {Marangos}},\ }\href@noop {} {\bibfield  {journal} {\bibinfo  {journal}
  {Contemp. Phys.}\ }\textbf {\bibinfo {volume} {52}},\ \bibinfo {pages} {551}
  (\bibinfo {year} {2011})}\BibitemShut {NoStop}%
\bibitem [{\citenamefont {Pellegrini}\ \emph {et~al.}(2016)\citenamefont
  {Pellegrini}, \citenamefont {Marinelli},\ and\ \citenamefont
  {Reiche}}]{Pellegrini16}%
  \BibitemOpen
  \bibfield  {author} {\bibinfo {author} {\bibfnamefont {C.}~\bibnamefont
  {Pellegrini}}, \bibinfo {author} {\bibfnamefont {A.}~\bibnamefont
  {Marinelli}}, \ and\ \bibinfo {author} {\bibfnamefont {S.}~\bibnamefont
  {Reiche}},\ }\href {\doibase 10.1103/RevModPhys.88.015006} {\bibfield
  {journal} {\bibinfo  {journal} {Rev. Mod. Phys.}\ }\textbf {\bibinfo {volume}
  {88}},\ \bibinfo {pages} {015006} (\bibinfo {year} {2016})}\BibitemShut
  {NoStop}%
\bibitem [{\citenamefont {Bostedt}\ \emph {et~al.}(2016)\citenamefont
  {Bostedt}, \citenamefont {Boutet}, \citenamefont {Fritz}, \citenamefont
  {Huang}, \citenamefont {Lee}, \citenamefont {Lemke}, \citenamefont {Robert},
  \citenamefont {Schlotter}, \citenamefont {Turner},\ and\ \citenamefont
  {Williams}}]{Bostedt16}%
  \BibitemOpen
  \bibfield  {author} {\bibinfo {author} {\bibfnamefont {C.}~\bibnamefont
  {Bostedt}}, \bibinfo {author} {\bibfnamefont {S.}~\bibnamefont {Boutet}},
  \bibinfo {author} {\bibfnamefont {D.~M.}\ \bibnamefont {Fritz}}, \bibinfo
  {author} {\bibfnamefont {Z.}~\bibnamefont {Huang}}, \bibinfo {author}
  {\bibfnamefont {H.~J.}\ \bibnamefont {Lee}}, \bibinfo {author} {\bibfnamefont
  {H.~T.}\ \bibnamefont {Lemke}}, \bibinfo {author} {\bibfnamefont
  {A.}~\bibnamefont {Robert}}, \bibinfo {author} {\bibfnamefont {W.~F.}\
  \bibnamefont {Schlotter}}, \bibinfo {author} {\bibfnamefont {J.~J.}\
  \bibnamefont {Turner}}, \ and\ \bibinfo {author} {\bibfnamefont {G.~J.}\
  \bibnamefont {Williams}},\ }\href {\doibase 10.1103/RevModPhys.88.015007}
  {\bibfield  {journal} {\bibinfo  {journal} {Rev. Mod. Phys.}\ }\textbf
  {\bibinfo {volume} {88}},\ \bibinfo {pages} {015007} (\bibinfo {year}
  {2016})}\BibitemShut {NoStop}%
\bibitem [{\citenamefont {Mimura}\ \emph {et~al.}(2014)\citenamefont {Mimura},
  \citenamefont {Yumoto}, \citenamefont {Matsuyama}, \citenamefont {Koyama},
  \citenamefont {Tono}, \citenamefont {Inubushi}, \citenamefont {Togashi},
  \citenamefont {Sato}, \citenamefont {Kim}, \citenamefont {Fukui},
  \citenamefont {Sano}, \citenamefont {Yabashi}, \citenamefont {Ohashi},
  \citenamefont {Ishikawa},\ and\ \citenamefont {Yamauchi}}]{Mimura14}%
  \BibitemOpen
  \bibfield  {author} {\bibinfo {author} {\bibfnamefont {H.}~\bibnamefont
  {Mimura}}, \bibinfo {author} {\bibfnamefont {H.}~\bibnamefont {Yumoto}},
  \bibinfo {author} {\bibfnamefont {S.}~\bibnamefont {Matsuyama}}, \bibinfo
  {author} {\bibfnamefont {T.}~\bibnamefont {Koyama}}, \bibinfo {author}
  {\bibfnamefont {K.}~\bibnamefont {Tono}}, \bibinfo {author} {\bibfnamefont
  {Y.}~\bibnamefont {Inubushi}}, \bibinfo {author} {\bibfnamefont
  {T.}~\bibnamefont {Togashi}}, \bibinfo {author} {\bibfnamefont
  {T.}~\bibnamefont {Sato}}, \bibinfo {author} {\bibfnamefont {J.}~\bibnamefont
  {Kim}}, \bibinfo {author} {\bibfnamefont {R.}~\bibnamefont {Fukui}}, \bibinfo
  {author} {\bibfnamefont {Y.}~\bibnamefont {Sano}}, \bibinfo {author}
  {\bibfnamefont {M.}~\bibnamefont {Yabashi}}, \bibinfo {author} {\bibfnamefont
  {H.}~\bibnamefont {Ohashi}}, \bibinfo {author} {\bibfnamefont
  {T.}~\bibnamefont {Ishikawa}}, \ and\ \bibinfo {author} {\bibfnamefont
  {K.}~\bibnamefont {Yamauchi}},\ }\href {\doibase 10.1038/ncomms4539}
  {\bibfield  {journal} {\bibinfo  {journal} {Nat. Commun.}\ }\textbf {\bibinfo
  {volume} {5}},\ \bibinfo {pages} {3539} (\bibinfo {year} {2014})}\BibitemShut
  {NoStop}%
\bibitem [{\citenamefont {Young}\ \emph {et~al.}(2010)\citenamefont {Young},
  \citenamefont {Kanter}, \citenamefont {Kr{\"a}ssig}, \citenamefont {Li},
  \citenamefont {March}, \citenamefont {Pratt}, \citenamefont {Santra},
  \citenamefont {Southworth}, \citenamefont {Rohringer}, \citenamefont
  {DiMauro}, \citenamefont {Doumy}, \citenamefont {Roedig}, \citenamefont
  {Berrah}, \citenamefont {Fang}, \citenamefont {Hoener}, \citenamefont
  {Bucksbaum}, \citenamefont {Cryan}, \citenamefont {Ghimire}, \citenamefont
  {Glownia}, \citenamefont {Reis}, \citenamefont {Bozek}, \citenamefont
  {Bostedt},\ and\ \citenamefont {Messerschmidt}}]{Young10}%
  \BibitemOpen
  \bibfield  {author} {\bibinfo {author} {\bibfnamefont {L.}~\bibnamefont
  {Young}}, \bibinfo {author} {\bibfnamefont {E.~P.}\ \bibnamefont {Kanter}},
  \bibinfo {author} {\bibfnamefont {B.}~\bibnamefont {Kr{\"a}ssig}}, \bibinfo
  {author} {\bibfnamefont {Y.}~\bibnamefont {Li}}, \bibinfo {author}
  {\bibfnamefont {A.~M.}\ \bibnamefont {March}}, \bibinfo {author}
  {\bibfnamefont {S.~T.}\ \bibnamefont {Pratt}}, \bibinfo {author}
  {\bibfnamefont {R.}~\bibnamefont {Santra}}, \bibinfo {author} {\bibfnamefont
  {S.~H.}\ \bibnamefont {Southworth}}, \bibinfo {author} {\bibfnamefont
  {N.}~\bibnamefont {Rohringer}}, \bibinfo {author} {\bibfnamefont {L.~F.}\
  \bibnamefont {DiMauro}}, \bibinfo {author} {\bibfnamefont {G.}~\bibnamefont
  {Doumy}}, \bibinfo {author} {\bibfnamefont {C.~A.}\ \bibnamefont {Roedig}},
  \bibinfo {author} {\bibfnamefont {N.}~\bibnamefont {Berrah}}, \bibinfo
  {author} {\bibfnamefont {L.}~\bibnamefont {Fang}}, \bibinfo {author}
  {\bibfnamefont {M.}~\bibnamefont {Hoener}}, \bibinfo {author} {\bibfnamefont
  {P.~H.}\ \bibnamefont {Bucksbaum}}, \bibinfo {author} {\bibfnamefont {J.~P.}\
  \bibnamefont {Cryan}}, \bibinfo {author} {\bibfnamefont {S.}~\bibnamefont
  {Ghimire}}, \bibinfo {author} {\bibfnamefont {J.~M.}\ \bibnamefont
  {Glownia}}, \bibinfo {author} {\bibfnamefont {D.~A.}\ \bibnamefont {Reis}},
  \bibinfo {author} {\bibfnamefont {J.~D.}\ \bibnamefont {Bozek}}, \bibinfo
  {author} {\bibfnamefont {C.}~\bibnamefont {Bostedt}}, \ and\ \bibinfo
  {author} {\bibfnamefont {M.}~\bibnamefont {Messerschmidt}},\ }\href@noop {}
  {\bibfield  {journal} {\bibinfo  {journal} {Nature}\ }\textbf {\bibinfo
  {volume} {466}},\ \bibinfo {pages} {56} (\bibinfo {year} {2010})}\BibitemShut
  {NoStop}%
\bibitem [{\citenamefont {Doumy}\ \emph {et~al.}(2011)\citenamefont {Doumy},
  \citenamefont {Roedig}, \citenamefont {Son}, \citenamefont {Blaga},
  \citenamefont {DiChiara}, \citenamefont {Santra}, \citenamefont {Berrah},
  \citenamefont {Bostedt}, \citenamefont {Bozek}, \citenamefont {Bucksbaum},
  \citenamefont {Cryan}, \citenamefont {Fang}, \citenamefont {Ghimire},
  \citenamefont {Glownia}, \citenamefont {Hoener}, \citenamefont {Kanter},
  \citenamefont {Kr{\"a}ssig}, \citenamefont {Kuebel}, \citenamefont
  {Messerschmidt}, \citenamefont {Paulus}, \citenamefont {Reis}, \citenamefont
  {Rohringer}, \citenamefont {Young}, \citenamefont {Agostini},\ and\
  \citenamefont {DiMauro}}]{Doumy11}%
  \BibitemOpen
  \bibfield  {author} {\bibinfo {author} {\bibfnamefont {G.}~\bibnamefont
  {Doumy}}, \bibinfo {author} {\bibfnamefont {C.}~\bibnamefont {Roedig}},
  \bibinfo {author} {\bibfnamefont {S.-K.}\ \bibnamefont {Son}}, \bibinfo
  {author} {\bibfnamefont {C.~I.}\ \bibnamefont {Blaga}}, \bibinfo {author}
  {\bibfnamefont {A.~D.}\ \bibnamefont {DiChiara}}, \bibinfo {author}
  {\bibfnamefont {R.}~\bibnamefont {Santra}}, \bibinfo {author} {\bibfnamefont
  {N.}~\bibnamefont {Berrah}}, \bibinfo {author} {\bibfnamefont
  {C.}~\bibnamefont {Bostedt}}, \bibinfo {author} {\bibfnamefont {J.~D.}\
  \bibnamefont {Bozek}}, \bibinfo {author} {\bibfnamefont {P.~H.}\ \bibnamefont
  {Bucksbaum}}, \bibinfo {author} {\bibfnamefont {J.~P.}\ \bibnamefont
  {Cryan}}, \bibinfo {author} {\bibfnamefont {L.}~\bibnamefont {Fang}},
  \bibinfo {author} {\bibfnamefont {S.}~\bibnamefont {Ghimire}}, \bibinfo
  {author} {\bibfnamefont {J.~M.}\ \bibnamefont {Glownia}}, \bibinfo {author}
  {\bibfnamefont {M.}~\bibnamefont {Hoener}}, \bibinfo {author} {\bibfnamefont
  {E.~P.}\ \bibnamefont {Kanter}}, \bibinfo {author} {\bibfnamefont
  {B.}~\bibnamefont {Kr{\"a}ssig}}, \bibinfo {author} {\bibfnamefont
  {M.}~\bibnamefont {Kuebel}}, \bibinfo {author} {\bibfnamefont
  {M.}~\bibnamefont {Messerschmidt}}, \bibinfo {author} {\bibfnamefont {G.~G.}\
  \bibnamefont {Paulus}}, \bibinfo {author} {\bibfnamefont {D.~A.}\
  \bibnamefont {Reis}}, \bibinfo {author} {\bibfnamefont {N.}~\bibnamefont
  {Rohringer}}, \bibinfo {author} {\bibfnamefont {L.}~\bibnamefont {Young}},
  \bibinfo {author} {\bibfnamefont {P.}~\bibnamefont {Agostini}}, \ and\
  \bibinfo {author} {\bibfnamefont {L.~F.}\ \bibnamefont {DiMauro}},\ }\href
  {\doibase 10.1103/PhysRevLett.106.083002} {\bibfield  {journal} {\bibinfo
  {journal} {Phys. Rev. Lett.}\ }\textbf {\bibinfo {volume} {106}},\ \bibinfo
  {pages} {083002} (\bibinfo {year} {2011})}\BibitemShut {NoStop}%
\bibitem [{\citenamefont {Rudek}\ \emph {et~al.}(2012)\citenamefont {Rudek},
  \citenamefont {Son}, \citenamefont {Foucar}, \citenamefont {Epp},
  \citenamefont {Erk}, \citenamefont {Hartmann}, \citenamefont {Adolph},
  \citenamefont {Andritschke}, \citenamefont {Aquila}, \citenamefont {Berrah},
  \citenamefont {Bostedt}, \citenamefont {Bozek}, \citenamefont {Coppola},
  \citenamefont {Filsinger}, \citenamefont {Gorke}, \citenamefont {Gorkhover},
  \citenamefont {Graafsma}, \citenamefont {Gumprecht}, \citenamefont
  {Hartmann}, \citenamefont {Hauser}, \citenamefont {Herrmann}, \citenamefont
  {Hirsemann}, \citenamefont {Holl}, \citenamefont {H{\"o}mke}, \citenamefont
  {Journel}, \citenamefont {Kaiser}, \citenamefont {Kimmel}, \citenamefont
  {Krasniqi}, \citenamefont {K{\"u}hnel}, \citenamefont {Matysek},
  \citenamefont {Messerschmidt}, \citenamefont {Miesner}, \citenamefont
  {M{\"o}ller}, \citenamefont {Moshammer}, \citenamefont {Nagaya},
  \citenamefont {Nilsson}, \citenamefont {Potdevin}, \citenamefont
  {Pietschner}, \citenamefont {Reich}, \citenamefont {Rupp}, \citenamefont
  {Schaller}, \citenamefont {Schlichting}, \citenamefont {Schmidt},
  \citenamefont {Schopper}, \citenamefont {Schorb}, \citenamefont
  {Schr{\"o}ter}, \citenamefont {Schulz}, \citenamefont {Simon}, \citenamefont
  {Soltau}, \citenamefont {Str{\"u}der}, \citenamefont {Ueda}, \citenamefont
  {Weidenspointner}, \citenamefont {Santra}, \citenamefont {Ullrich},
  \citenamefont {Rudenko},\ and\ \citenamefont {Rolles}}]{Rudek12}%
  \BibitemOpen
  \bibfield  {author} {\bibinfo {author} {\bibfnamefont {B.}~\bibnamefont
  {Rudek}}, \bibinfo {author} {\bibfnamefont {S.-K.}\ \bibnamefont {Son}},
  \bibinfo {author} {\bibfnamefont {L.}~\bibnamefont {Foucar}}, \bibinfo
  {author} {\bibfnamefont {S.~W.}\ \bibnamefont {Epp}}, \bibinfo {author}
  {\bibfnamefont {B.}~\bibnamefont {Erk}}, \bibinfo {author} {\bibfnamefont
  {R.}~\bibnamefont {Hartmann}}, \bibinfo {author} {\bibfnamefont
  {M.}~\bibnamefont {Adolph}}, \bibinfo {author} {\bibfnamefont
  {R.}~\bibnamefont {Andritschke}}, \bibinfo {author} {\bibfnamefont
  {A.}~\bibnamefont {Aquila}}, \bibinfo {author} {\bibfnamefont
  {N.}~\bibnamefont {Berrah}}, \bibinfo {author} {\bibfnamefont
  {C.}~\bibnamefont {Bostedt}}, \bibinfo {author} {\bibfnamefont
  {J.}~\bibnamefont {Bozek}}, \bibinfo {author} {\bibfnamefont
  {N.}~\bibnamefont {Coppola}}, \bibinfo {author} {\bibfnamefont
  {F.}~\bibnamefont {Filsinger}}, \bibinfo {author} {\bibfnamefont
  {H.}~\bibnamefont {Gorke}}, \bibinfo {author} {\bibfnamefont
  {T.}~\bibnamefont {Gorkhover}}, \bibinfo {author} {\bibfnamefont
  {H.}~\bibnamefont {Graafsma}}, \bibinfo {author} {\bibfnamefont
  {L.}~\bibnamefont {Gumprecht}}, \bibinfo {author} {\bibfnamefont
  {A.}~\bibnamefont {Hartmann}}, \bibinfo {author} {\bibfnamefont
  {G.}~\bibnamefont {Hauser}}, \bibinfo {author} {\bibfnamefont
  {S.}~\bibnamefont {Herrmann}}, \bibinfo {author} {\bibfnamefont
  {H.}~\bibnamefont {Hirsemann}}, \bibinfo {author} {\bibfnamefont
  {P.}~\bibnamefont {Holl}}, \bibinfo {author} {\bibfnamefont {A.}~\bibnamefont
  {H{\"o}mke}}, \bibinfo {author} {\bibfnamefont {L.}~\bibnamefont {Journel}},
  \bibinfo {author} {\bibfnamefont {C.}~\bibnamefont {Kaiser}}, \bibinfo
  {author} {\bibfnamefont {N.}~\bibnamefont {Kimmel}}, \bibinfo {author}
  {\bibfnamefont {F.}~\bibnamefont {Krasniqi}}, \bibinfo {author}
  {\bibfnamefont {K.-U.}\ \bibnamefont {K{\"u}hnel}}, \bibinfo {author}
  {\bibfnamefont {M.}~\bibnamefont {Matysek}}, \bibinfo {author} {\bibfnamefont
  {M.}~\bibnamefont {Messerschmidt}}, \bibinfo {author} {\bibfnamefont
  {D.}~\bibnamefont {Miesner}}, \bibinfo {author} {\bibfnamefont
  {T.}~\bibnamefont {M{\"o}ller}}, \bibinfo {author} {\bibfnamefont
  {R.}~\bibnamefont {Moshammer}}, \bibinfo {author} {\bibfnamefont
  {K.}~\bibnamefont {Nagaya}}, \bibinfo {author} {\bibfnamefont
  {B.}~\bibnamefont {Nilsson}}, \bibinfo {author} {\bibfnamefont
  {G.}~\bibnamefont {Potdevin}}, \bibinfo {author} {\bibfnamefont
  {D.}~\bibnamefont {Pietschner}}, \bibinfo {author} {\bibfnamefont
  {C.}~\bibnamefont {Reich}}, \bibinfo {author} {\bibfnamefont
  {D.}~\bibnamefont {Rupp}}, \bibinfo {author} {\bibfnamefont {G.}~\bibnamefont
  {Schaller}}, \bibinfo {author} {\bibfnamefont {I.}~\bibnamefont
  {Schlichting}}, \bibinfo {author} {\bibfnamefont {C.}~\bibnamefont
  {Schmidt}}, \bibinfo {author} {\bibfnamefont {F.}~\bibnamefont {Schopper}},
  \bibinfo {author} {\bibfnamefont {S.}~\bibnamefont {Schorb}}, \bibinfo
  {author} {\bibfnamefont {C.-D.}\ \bibnamefont {Schr{\"o}ter}}, \bibinfo
  {author} {\bibfnamefont {J.}~\bibnamefont {Schulz}}, \bibinfo {author}
  {\bibfnamefont {M.}~\bibnamefont {Simon}}, \bibinfo {author} {\bibfnamefont
  {H.}~\bibnamefont {Soltau}}, \bibinfo {author} {\bibfnamefont
  {L.}~\bibnamefont {Str{\"u}der}}, \bibinfo {author} {\bibfnamefont
  {K.}~\bibnamefont {Ueda}}, \bibinfo {author} {\bibfnamefont {G.}~\bibnamefont
  {Weidenspointner}}, \bibinfo {author} {\bibfnamefont {R.}~\bibnamefont
  {Santra}}, \bibinfo {author} {\bibfnamefont {J.}~\bibnamefont {Ullrich}},
  \bibinfo {author} {\bibfnamefont {A.}~\bibnamefont {Rudenko}}, \ and\
  \bibinfo {author} {\bibfnamefont {D.}~\bibnamefont {Rolles}},\ }\href
  {\doibase 10.1038/nphoton.2012.261} {\bibfield  {journal} {\bibinfo
  {journal} {Nature Photon.}\ }\textbf {\bibinfo {volume} {6}},\ \bibinfo
  {pages} {858} (\bibinfo {year} {2012})}\BibitemShut {NoStop}%
\bibitem [{\citenamefont {Fukuzawa}\ \emph {et~al.}(2013)\citenamefont
  {Fukuzawa}, \citenamefont {Son}, \citenamefont {Motomura}, \citenamefont
  {Mondal}, \citenamefont {Nagaya}, \citenamefont {Wada}, \citenamefont {Liu},
  \citenamefont {Feifel}, \citenamefont {Tachibana}, \citenamefont {Ito},
  \citenamefont {Kimura}, \citenamefont {Sakai}, \citenamefont {Matsunami},
  \citenamefont {Hayashita}, \citenamefont {Kajikawa}, \citenamefont
  {Johnsson}, \citenamefont {Siano}, \citenamefont {Kukk}, \citenamefont
  {Rudek}, \citenamefont {Erk}, \citenamefont {Foucar}, \citenamefont {Robert},
  \citenamefont {Miron}, \citenamefont {Tono}, \citenamefont {Inubushi},
  \citenamefont {Hatsui}, \citenamefont {Yabashi}, \citenamefont {Yao},
  \citenamefont {Santra},\ and\ \citenamefont {Ueda}}]{Fukuzawa13}%
  \BibitemOpen
  \bibfield  {author} {\bibinfo {author} {\bibfnamefont {H.}~\bibnamefont
  {Fukuzawa}}, \bibinfo {author} {\bibfnamefont {S.-K.}\ \bibnamefont {Son}},
  \bibinfo {author} {\bibfnamefont {K.}~\bibnamefont {Motomura}}, \bibinfo
  {author} {\bibfnamefont {S.}~\bibnamefont {Mondal}}, \bibinfo {author}
  {\bibfnamefont {K.}~\bibnamefont {Nagaya}}, \bibinfo {author} {\bibfnamefont
  {S.}~\bibnamefont {Wada}}, \bibinfo {author} {\bibfnamefont {X.-J.}\
  \bibnamefont {Liu}}, \bibinfo {author} {\bibfnamefont {R.}~\bibnamefont
  {Feifel}}, \bibinfo {author} {\bibfnamefont {T.}~\bibnamefont {Tachibana}},
  \bibinfo {author} {\bibfnamefont {Y.}~\bibnamefont {Ito}}, \bibinfo {author}
  {\bibfnamefont {M.}~\bibnamefont {Kimura}}, \bibinfo {author} {\bibfnamefont
  {T.}~\bibnamefont {Sakai}}, \bibinfo {author} {\bibfnamefont
  {K.}~\bibnamefont {Matsunami}}, \bibinfo {author} {\bibfnamefont
  {H.}~\bibnamefont {Hayashita}}, \bibinfo {author} {\bibfnamefont
  {J.}~\bibnamefont {Kajikawa}}, \bibinfo {author} {\bibfnamefont
  {P.}~\bibnamefont {Johnsson}}, \bibinfo {author} {\bibfnamefont
  {M.}~\bibnamefont {Siano}}, \bibinfo {author} {\bibfnamefont
  {E.}~\bibnamefont {Kukk}}, \bibinfo {author} {\bibfnamefont {B.}~\bibnamefont
  {Rudek}}, \bibinfo {author} {\bibfnamefont {B.}~\bibnamefont {Erk}}, \bibinfo
  {author} {\bibfnamefont {L.}~\bibnamefont {Foucar}}, \bibinfo {author}
  {\bibfnamefont {E.}~\bibnamefont {Robert}}, \bibinfo {author} {\bibfnamefont
  {C.}~\bibnamefont {Miron}}, \bibinfo {author} {\bibfnamefont
  {K.}~\bibnamefont {Tono}}, \bibinfo {author} {\bibfnamefont {Y.}~\bibnamefont
  {Inubushi}}, \bibinfo {author} {\bibfnamefont {T.}~\bibnamefont {Hatsui}},
  \bibinfo {author} {\bibfnamefont {M.}~\bibnamefont {Yabashi}}, \bibinfo
  {author} {\bibfnamefont {M.}~\bibnamefont {Yao}}, \bibinfo {author}
  {\bibfnamefont {R.}~\bibnamefont {Santra}}, \ and\ \bibinfo {author}
  {\bibfnamefont {K.}~\bibnamefont {Ueda}},\ }\href {\doibase
  10.1103/PhysRevLett.110.173005} {\bibfield  {journal} {\bibinfo  {journal}
  {Phys. Rev. Lett.}\ }\textbf {\bibinfo {volume} {110}},\ \bibinfo {pages}
  {173005} (\bibinfo {year} {2013})}\BibitemShut {NoStop}%
\bibitem [{\citenamefont {Rudek}\ \emph {et~al.}(2018)\citenamefont {Rudek},
  \citenamefont {Toyota}, \citenamefont {Foucar}, \citenamefont {Erk},
  \citenamefont {Boll}, \citenamefont {Bomme}, \citenamefont {Correa},
  \citenamefont {Carron}, \citenamefont {Boutet}, \citenamefont {Williams},
  \citenamefont {Ferguson}, \citenamefont {Alonso-Mori}, \citenamefont
  {Koglin}, \citenamefont {Gorkhover}, \citenamefont {Bucher}, \citenamefont
  {Lehmann}, \citenamefont {Kr{"a}ssig}, \citenamefont {Southworth},
  \citenamefont {Young}, \citenamefont {Bostedt}, \citenamefont {Ueda},
  \citenamefont {Marchenko}, \citenamefont {Simon}, \citenamefont {Jurek},
  \citenamefont {Santra}, \citenamefont {Rudenko}, \citenamefont {Son},\ and\
  \citenamefont {Rolles}}]{Rudek18}%
  \BibitemOpen
  \bibfield  {author} {\bibinfo {author} {\bibfnamefont {B.}~\bibnamefont
  {Rudek}}, \bibinfo {author} {\bibfnamefont {K.}~\bibnamefont {Toyota}},
  \bibinfo {author} {\bibfnamefont {L.}~\bibnamefont {Foucar}}, \bibinfo
  {author} {\bibfnamefont {B.}~\bibnamefont {Erk}}, \bibinfo {author}
  {\bibfnamefont {R.}~\bibnamefont {Boll}}, \bibinfo {author} {\bibfnamefont
  {C.}~\bibnamefont {Bomme}}, \bibinfo {author} {\bibfnamefont
  {J.}~\bibnamefont {Correa}}, \bibinfo {author} {\bibfnamefont
  {S.}~\bibnamefont {Carron}}, \bibinfo {author} {\bibfnamefont
  {S.}~\bibnamefont {Boutet}}, \bibinfo {author} {\bibfnamefont {G.~J.}\
  \bibnamefont {Williams}}, \bibinfo {author} {\bibfnamefont {K.~R.}\
  \bibnamefont {Ferguson}}, \bibinfo {author} {\bibfnamefont {R.}~\bibnamefont
  {Alonso-Mori}}, \bibinfo {author} {\bibfnamefont {J.~E.}\ \bibnamefont
  {Koglin}}, \bibinfo {author} {\bibfnamefont {T.}~\bibnamefont {Gorkhover}},
  \bibinfo {author} {\bibfnamefont {M.}~\bibnamefont {Bucher}}, \bibinfo
  {author} {\bibfnamefont {C.~S.}\ \bibnamefont {Lehmann}}, \bibinfo {author}
  {\bibfnamefont {B.}~\bibnamefont {Kr{"a}ssig}}, \bibinfo {author}
  {\bibfnamefont {S.~H.}\ \bibnamefont {Southworth}}, \bibinfo {author}
  {\bibfnamefont {L.}~\bibnamefont {Young}}, \bibinfo {author} {\bibfnamefont
  {C.}~\bibnamefont {Bostedt}}, \bibinfo {author} {\bibfnamefont
  {K.}~\bibnamefont {Ueda}}, \bibinfo {author} {\bibfnamefont {T.}~\bibnamefont
  {Marchenko}}, \bibinfo {author} {\bibfnamefont {M.}~\bibnamefont {Simon}},
  \bibinfo {author} {\bibfnamefont {Z.}~\bibnamefont {Jurek}}, \bibinfo
  {author} {\bibfnamefont {R.}~\bibnamefont {Santra}}, \bibinfo {author}
  {\bibfnamefont {A.}~\bibnamefont {Rudenko}}, \bibinfo {author} {\bibfnamefont
  {S.-K.}\ \bibnamefont {Son}}, \ and\ \bibinfo {author} {\bibfnamefont
  {D.}~\bibnamefont {Rolles}},\ }\href {\doibase 10.1038/s41467-018-06745-6}
  {\bibfield  {journal} {\bibinfo  {journal} {Nat. Commun.}\ }\textbf {\bibinfo
  {volume} {9}},\ \bibinfo {pages} {4200} (\bibinfo {year} {2018})}\BibitemShut
  {NoStop}%
\bibitem [{\citenamefont {Hoener}\ \emph {et~al.}(2010)\citenamefont {Hoener},
  \citenamefont {Fang}, \citenamefont {Kornilov}, \citenamefont {Gessner},
  \citenamefont {Pratt}, \citenamefont {G\"uhr}, \citenamefont {Kanter},
  \citenamefont {Blaga}, \citenamefont {Bostedt}, \citenamefont {Bozek},
  \citenamefont {Bucksbaum}, \citenamefont {Buth}, \citenamefont {Chen},
  \citenamefont {Coffee}, \citenamefont {Cryan}, \citenamefont {DiMauro},
  \citenamefont {Glownia}, \citenamefont {Hosler}, \citenamefont {Kukk},
  \citenamefont {Leone}, \citenamefont {McFarland}, \citenamefont
  {Messerschmidt}, \citenamefont {Murphy}, \citenamefont {Petrovic},
  \citenamefont {Rolles},\ and\ \citenamefont {Berrah}}]{Hoener10}%
  \BibitemOpen
  \bibfield  {author} {\bibinfo {author} {\bibfnamefont {M.}~\bibnamefont
  {Hoener}}, \bibinfo {author} {\bibfnamefont {L.}~\bibnamefont {Fang}},
  \bibinfo {author} {\bibfnamefont {O.}~\bibnamefont {Kornilov}}, \bibinfo
  {author} {\bibfnamefont {O.}~\bibnamefont {Gessner}}, \bibinfo {author}
  {\bibfnamefont {S.~T.}\ \bibnamefont {Pratt}}, \bibinfo {author}
  {\bibfnamefont {M.}~\bibnamefont {G\"uhr}}, \bibinfo {author} {\bibfnamefont
  {E.~P.}\ \bibnamefont {Kanter}}, \bibinfo {author} {\bibfnamefont
  {C.}~\bibnamefont {Blaga}}, \bibinfo {author} {\bibfnamefont
  {C.}~\bibnamefont {Bostedt}}, \bibinfo {author} {\bibfnamefont {J.~D.}\
  \bibnamefont {Bozek}}, \bibinfo {author} {\bibfnamefont {P.~H.}\ \bibnamefont
  {Bucksbaum}}, \bibinfo {author} {\bibfnamefont {C.}~\bibnamefont {Buth}},
  \bibinfo {author} {\bibfnamefont {M.}~\bibnamefont {Chen}}, \bibinfo {author}
  {\bibfnamefont {R.}~\bibnamefont {Coffee}}, \bibinfo {author} {\bibfnamefont
  {J.}~\bibnamefont {Cryan}}, \bibinfo {author} {\bibfnamefont
  {L.}~\bibnamefont {DiMauro}}, \bibinfo {author} {\bibfnamefont
  {M.}~\bibnamefont {Glownia}}, \bibinfo {author} {\bibfnamefont
  {E.}~\bibnamefont {Hosler}}, \bibinfo {author} {\bibfnamefont
  {E.}~\bibnamefont {Kukk}}, \bibinfo {author} {\bibfnamefont {S.~R.}\
  \bibnamefont {Leone}}, \bibinfo {author} {\bibfnamefont {B.}~\bibnamefont
  {McFarland}}, \bibinfo {author} {\bibfnamefont {M.}~\bibnamefont
  {Messerschmidt}}, \bibinfo {author} {\bibfnamefont {B.}~\bibnamefont
  {Murphy}}, \bibinfo {author} {\bibfnamefont {V.}~\bibnamefont {Petrovic}},
  \bibinfo {author} {\bibfnamefont {D.}~\bibnamefont {Rolles}}, \ and\ \bibinfo
  {author} {\bibfnamefont {N.}~\bibnamefont {Berrah}},\ }\href {\doibase
  10.1103/PhysRevLett.104.253002} {\bibfield  {journal} {\bibinfo  {journal}
  {Phys. Rev. Lett.}\ }\textbf {\bibinfo {volume} {104}},\ \bibinfo {pages}
  {253002} (\bibinfo {year} {2010})}\BibitemShut {NoStop}%
\bibitem [{\citenamefont {Fang}\ \emph {et~al.}(2010)\citenamefont {Fang},
  \citenamefont {Hoener}, \citenamefont {Gessner}, \citenamefont {Tarantelli},
  \citenamefont {Pratt}, \citenamefont {Kornilov}, \citenamefont {Buth},
  \citenamefont {G\"uhr}, \citenamefont {Kanter}, \citenamefont {Bostedt},
  \citenamefont {Bozek}, \citenamefont {Bucksbaum}, \citenamefont {Chen},
  \citenamefont {Coffee}, \citenamefont {Cryan}, \citenamefont {Glownia},
  \citenamefont {Kukk}, \citenamefont {Leone},\ and\ \citenamefont
  {Berrah}}]{Fang10}%
  \BibitemOpen
  \bibfield  {author} {\bibinfo {author} {\bibfnamefont {L.}~\bibnamefont
  {Fang}}, \bibinfo {author} {\bibfnamefont {M.}~\bibnamefont {Hoener}},
  \bibinfo {author} {\bibfnamefont {O.}~\bibnamefont {Gessner}}, \bibinfo
  {author} {\bibfnamefont {F.}~\bibnamefont {Tarantelli}}, \bibinfo {author}
  {\bibfnamefont {S.~T.}\ \bibnamefont {Pratt}}, \bibinfo {author}
  {\bibfnamefont {O.}~\bibnamefont {Kornilov}}, \bibinfo {author}
  {\bibfnamefont {C.}~\bibnamefont {Buth}}, \bibinfo {author} {\bibfnamefont
  {M.}~\bibnamefont {G\"uhr}}, \bibinfo {author} {\bibfnamefont {E.~P.}\
  \bibnamefont {Kanter}}, \bibinfo {author} {\bibfnamefont {C.}~\bibnamefont
  {Bostedt}}, \bibinfo {author} {\bibfnamefont {J.~D.}\ \bibnamefont {Bozek}},
  \bibinfo {author} {\bibfnamefont {P.~H.}\ \bibnamefont {Bucksbaum}}, \bibinfo
  {author} {\bibfnamefont {M.}~\bibnamefont {Chen}}, \bibinfo {author}
  {\bibfnamefont {R.}~\bibnamefont {Coffee}}, \bibinfo {author} {\bibfnamefont
  {J.}~\bibnamefont {Cryan}}, \bibinfo {author} {\bibfnamefont
  {M.}~\bibnamefont {Glownia}}, \bibinfo {author} {\bibfnamefont
  {E.}~\bibnamefont {Kukk}}, \bibinfo {author} {\bibfnamefont {S.~R.}\
  \bibnamefont {Leone}}, \ and\ \bibinfo {author} {\bibfnamefont
  {N.}~\bibnamefont {Berrah}},\ }\href {\doibase
  10.1103/PhysRevLett.105.083005} {\bibfield  {journal} {\bibinfo  {journal}
  {Phys. Rev. Lett.}\ }\textbf {\bibinfo {volume} {105}},\ \bibinfo {pages}
  {083005} (\bibinfo {year} {2010})}\BibitemShut {NoStop}%
\bibitem [{\citenamefont {Erk}\ \emph {et~al.}(2013{\natexlab{a}})\citenamefont
  {Erk}, \citenamefont {Rolles}, \citenamefont {Foucar}, \citenamefont {Rudek},
  \citenamefont {Epp}, \citenamefont {Cryle}, \citenamefont {Bostedt},
  \citenamefont {Schorb}, \citenamefont {Bozek}, \citenamefont {Rouzee},
  \citenamefont {Hundertmark}, \citenamefont {Marchenko}, \citenamefont
  {Simon}, \citenamefont {Filsinger}, \citenamefont {Christensen},
  \citenamefont {De}, \citenamefont {Trippel}, \citenamefont {K\"upper},
  \citenamefont {Stapelfeldt}, \citenamefont {Wada}, \citenamefont {Ueda},
  \citenamefont {Swiggers}, \citenamefont {Messerschmidt}, \citenamefont
  {Schr\"oter}, \citenamefont {Moshammer}, \citenamefont {Schlichting},
  \citenamefont {Ullrich},\ and\ \citenamefont {Rudenko}}]{Erk13}%
  \BibitemOpen
  \bibfield  {author} {\bibinfo {author} {\bibfnamefont {B.}~\bibnamefont
  {Erk}}, \bibinfo {author} {\bibfnamefont {D.}~\bibnamefont {Rolles}},
  \bibinfo {author} {\bibfnamefont {L.}~\bibnamefont {Foucar}}, \bibinfo
  {author} {\bibfnamefont {B.}~\bibnamefont {Rudek}}, \bibinfo {author}
  {\bibfnamefont {S.~W.}\ \bibnamefont {Epp}}, \bibinfo {author} {\bibfnamefont
  {M.}~\bibnamefont {Cryle}}, \bibinfo {author} {\bibfnamefont
  {C.}~\bibnamefont {Bostedt}}, \bibinfo {author} {\bibfnamefont
  {S.}~\bibnamefont {Schorb}}, \bibinfo {author} {\bibfnamefont
  {J.}~\bibnamefont {Bozek}}, \bibinfo {author} {\bibfnamefont
  {A.}~\bibnamefont {Rouzee}}, \bibinfo {author} {\bibfnamefont
  {A.}~\bibnamefont {Hundertmark}}, \bibinfo {author} {\bibfnamefont
  {T.}~\bibnamefont {Marchenko}}, \bibinfo {author} {\bibfnamefont
  {M.}~\bibnamefont {Simon}}, \bibinfo {author} {\bibfnamefont
  {F.}~\bibnamefont {Filsinger}}, \bibinfo {author} {\bibfnamefont
  {L.}~\bibnamefont {Christensen}}, \bibinfo {author} {\bibfnamefont
  {S.}~\bibnamefont {De}}, \bibinfo {author} {\bibfnamefont {S.}~\bibnamefont
  {Trippel}}, \bibinfo {author} {\bibfnamefont {J.}~\bibnamefont {K\"upper}},
  \bibinfo {author} {\bibfnamefont {H.}~\bibnamefont {Stapelfeldt}}, \bibinfo
  {author} {\bibfnamefont {S.}~\bibnamefont {Wada}}, \bibinfo {author}
  {\bibfnamefont {K.}~\bibnamefont {Ueda}}, \bibinfo {author} {\bibfnamefont
  {M.}~\bibnamefont {Swiggers}}, \bibinfo {author} {\bibfnamefont
  {M.}~\bibnamefont {Messerschmidt}}, \bibinfo {author} {\bibfnamefont {C.~D.}\
  \bibnamefont {Schr\"oter}}, \bibinfo {author} {\bibfnamefont
  {R.}~\bibnamefont {Moshammer}}, \bibinfo {author} {\bibfnamefont
  {I.}~\bibnamefont {Schlichting}}, \bibinfo {author} {\bibfnamefont
  {J.}~\bibnamefont {Ullrich}}, \ and\ \bibinfo {author} {\bibfnamefont
  {A.}~\bibnamefont {Rudenko}},\ }\href@noop {} {\bibfield  {journal} {\bibinfo
   {journal} {Phys. Rev. Lett.}\ }\textbf {\bibinfo {volume} {110}},\ \bibinfo
  {pages} {053003} (\bibinfo {year} {2013}{\natexlab{a}})}\BibitemShut
  {NoStop}%
\bibitem [{\citenamefont {Erk}\ \emph {et~al.}(2013{\natexlab{b}})\citenamefont
  {Erk}, \citenamefont {Rolles}, \citenamefont {Foucar}, \citenamefont {Rudek},
  \citenamefont {Epp}, \citenamefont {Cryle}, \citenamefont {Bostedt},
  \citenamefont {Schorb}, \citenamefont {Bozek}, \citenamefont {Rouzee},
  \citenamefont {Hundertmark}, \citenamefont {Marchenko}, \citenamefont
  {Simon}, \citenamefont {Filsinger}, \citenamefont {Christensen},
  \citenamefont {De}, \citenamefont {Trippel}, \citenamefont {K{\"u}pper},
  \citenamefont {Stapelfeldt}, \citenamefont {Wada}, \citenamefont {Ueda},
  \citenamefont {Swiggers}, \citenamefont {Messerschmidt}, \citenamefont
  {Schr{\"o}ter}, \citenamefont {Moshammer}, \citenamefont {Schlichting},
  \citenamefont {Ullrich},\ and\ \citenamefont {Rudenko}}]{Erk13a}%
  \BibitemOpen
  \bibfield  {author} {\bibinfo {author} {\bibfnamefont {B.}~\bibnamefont
  {Erk}}, \bibinfo {author} {\bibfnamefont {D.}~\bibnamefont {Rolles}},
  \bibinfo {author} {\bibfnamefont {L.}~\bibnamefont {Foucar}}, \bibinfo
  {author} {\bibfnamefont {B.}~\bibnamefont {Rudek}}, \bibinfo {author}
  {\bibfnamefont {S.~W.}\ \bibnamefont {Epp}}, \bibinfo {author} {\bibfnamefont
  {M.}~\bibnamefont {Cryle}}, \bibinfo {author} {\bibfnamefont
  {C.}~\bibnamefont {Bostedt}}, \bibinfo {author} {\bibfnamefont
  {S.}~\bibnamefont {Schorb}}, \bibinfo {author} {\bibfnamefont
  {J.}~\bibnamefont {Bozek}}, \bibinfo {author} {\bibfnamefont
  {A.}~\bibnamefont {Rouzee}}, \bibinfo {author} {\bibfnamefont
  {A.}~\bibnamefont {Hundertmark}}, \bibinfo {author} {\bibfnamefont
  {T.}~\bibnamefont {Marchenko}}, \bibinfo {author} {\bibfnamefont
  {M.}~\bibnamefont {Simon}}, \bibinfo {author} {\bibfnamefont
  {F.}~\bibnamefont {Filsinger}}, \bibinfo {author} {\bibfnamefont
  {L.}~\bibnamefont {Christensen}}, \bibinfo {author} {\bibfnamefont
  {S.}~\bibnamefont {De}}, \bibinfo {author} {\bibfnamefont {S.}~\bibnamefont
  {Trippel}}, \bibinfo {author} {\bibfnamefont {J.}~\bibnamefont {K{\"u}pper}},
  \bibinfo {author} {\bibfnamefont {H.}~\bibnamefont {Stapelfeldt}}, \bibinfo
  {author} {\bibfnamefont {S.}~\bibnamefont {Wada}}, \bibinfo {author}
  {\bibfnamefont {K.}~\bibnamefont {Ueda}}, \bibinfo {author} {\bibfnamefont
  {M.}~\bibnamefont {Swiggers}}, \bibinfo {author} {\bibfnamefont
  {M.}~\bibnamefont {Messerschmidt}}, \bibinfo {author} {\bibfnamefont {C.~D.}\
  \bibnamefont {Schr{\"o}ter}}, \bibinfo {author} {\bibfnamefont
  {R.}~\bibnamefont {Moshammer}}, \bibinfo {author} {\bibfnamefont
  {I.}~\bibnamefont {Schlichting}}, \bibinfo {author} {\bibfnamefont
  {J.}~\bibnamefont {Ullrich}}, \ and\ \bibinfo {author} {\bibfnamefont
  {A.}~\bibnamefont {Rudenko}},\ }\href {\doibase
  10.1088/0953-4075/46/16/164031} {\bibfield  {journal} {\bibinfo  {journal}
  {J. Phys. B: At. Mol. Opt. Phys.}\ }\textbf {\bibinfo {volume} {46}},\
  \bibinfo {pages} {164031} (\bibinfo {year} {2013}{\natexlab{b}})}\BibitemShut
  {NoStop}%
\bibitem [{\citenamefont {Boll}\ \emph {et~al.}(2016)\citenamefont {Boll},
  \citenamefont {Erk}, \citenamefont {Coffee}, \citenamefont {Trippel},
  \citenamefont {Kierspel}, \citenamefont {Bomme}, \citenamefont {Bozek},
  \citenamefont {Burkett}, \citenamefont {Carron}, \citenamefont {Ferguson},
  \citenamefont {Foucar}, \citenamefont {K{\"u}pper}, \citenamefont
  {Marchenko}, \citenamefont {Miron}, \citenamefont {Patanen}, \citenamefont
  {Osipov}, \citenamefont {Schorb}, \citenamefont {Simon}, \citenamefont
  {Swiggers}, \citenamefont {Techert}, \citenamefont {Ueda}, \citenamefont
  {Bostedt}, \citenamefont {Rolles},\ and\ \citenamefont {Rudenko}}]{Boll16}%
  \BibitemOpen
  \bibfield  {author} {\bibinfo {author} {\bibfnamefont {R.}~\bibnamefont
  {Boll}}, \bibinfo {author} {\bibfnamefont {B.}~\bibnamefont {Erk}}, \bibinfo
  {author} {\bibfnamefont {R.}~\bibnamefont {Coffee}}, \bibinfo {author}
  {\bibfnamefont {S.}~\bibnamefont {Trippel}}, \bibinfo {author} {\bibfnamefont
  {T.}~\bibnamefont {Kierspel}}, \bibinfo {author} {\bibfnamefont
  {C.}~\bibnamefont {Bomme}}, \bibinfo {author} {\bibfnamefont {J.~D.}\
  \bibnamefont {Bozek}}, \bibinfo {author} {\bibfnamefont {M.}~\bibnamefont
  {Burkett}}, \bibinfo {author} {\bibfnamefont {S.}~\bibnamefont {Carron}},
  \bibinfo {author} {\bibfnamefont {K.~R.}\ \bibnamefont {Ferguson}}, \bibinfo
  {author} {\bibfnamefont {L.}~\bibnamefont {Foucar}}, \bibinfo {author}
  {\bibfnamefont {J.}~\bibnamefont {K{\"u}pper}}, \bibinfo {author}
  {\bibfnamefont {T.}~\bibnamefont {Marchenko}}, \bibinfo {author}
  {\bibfnamefont {C.}~\bibnamefont {Miron}}, \bibinfo {author} {\bibfnamefont
  {M.}~\bibnamefont {Patanen}}, \bibinfo {author} {\bibfnamefont
  {T.}~\bibnamefont {Osipov}}, \bibinfo {author} {\bibfnamefont
  {S.}~\bibnamefont {Schorb}}, \bibinfo {author} {\bibfnamefont
  {M.}~\bibnamefont {Simon}}, \bibinfo {author} {\bibfnamefont
  {M.}~\bibnamefont {Swiggers}}, \bibinfo {author} {\bibfnamefont
  {S.}~\bibnamefont {Techert}}, \bibinfo {author} {\bibfnamefont
  {K.}~\bibnamefont {Ueda}}, \bibinfo {author} {\bibfnamefont {C.}~\bibnamefont
  {Bostedt}}, \bibinfo {author} {\bibfnamefont {D.}~\bibnamefont {Rolles}}, \
  and\ \bibinfo {author} {\bibfnamefont {A.}~\bibnamefont {Rudenko}},\ }\href
  {\doibase 10.1063/1.4944344} {\bibfield  {journal} {\bibinfo  {journal}
  {Struct. Dyn.}\ }\textbf {\bibinfo {volume} {3}},\ \bibinfo {pages} {043207}
  (\bibinfo {year} {2016})}\BibitemShut {NoStop}%
\bibitem [{\citenamefont {Motomura}\ \emph {et~al.}(2015)\citenamefont
  {Motomura}, \citenamefont {Kukk}, \citenamefont {Fukuzawa}, \citenamefont
  {Wada}, \citenamefont {Nagaya}, \citenamefont {Ohmura}, \citenamefont
  {Mondal}, \citenamefont {Tachibana}, \citenamefont {Ito}, \citenamefont
  {Koga}, \citenamefont {Sakai}, \citenamefont {Matsunami}, \citenamefont
  {Rudenko}, \citenamefont {Nicolas}, \citenamefont {Liu}, \citenamefont
  {Miron}, \citenamefont {Zhang}, \citenamefont {Jiang}, \citenamefont {Chen},
  \citenamefont {Anand}, \citenamefont {Kim}, \citenamefont {Tono},
  \citenamefont {Yabashi}, \citenamefont {Yao},\ and\ \citenamefont
  {Ueda}}]{Motomura15}%
  \BibitemOpen
  \bibfield  {author} {\bibinfo {author} {\bibfnamefont {K.}~\bibnamefont
  {Motomura}}, \bibinfo {author} {\bibfnamefont {E.}~\bibnamefont {Kukk}},
  \bibinfo {author} {\bibfnamefont {H.}~\bibnamefont {Fukuzawa}}, \bibinfo
  {author} {\bibfnamefont {S.}~\bibnamefont {Wada}}, \bibinfo {author}
  {\bibfnamefont {K.}~\bibnamefont {Nagaya}}, \bibinfo {author} {\bibfnamefont
  {S.}~\bibnamefont {Ohmura}}, \bibinfo {author} {\bibfnamefont
  {S.}~\bibnamefont {Mondal}}, \bibinfo {author} {\bibfnamefont
  {T.}~\bibnamefont {Tachibana}}, \bibinfo {author} {\bibfnamefont
  {Y.}~\bibnamefont {Ito}}, \bibinfo {author} {\bibfnamefont {R.}~\bibnamefont
  {Koga}}, \bibinfo {author} {\bibfnamefont {T.}~\bibnamefont {Sakai}},
  \bibinfo {author} {\bibfnamefont {K.}~\bibnamefont {Matsunami}}, \bibinfo
  {author} {\bibfnamefont {A.}~\bibnamefont {Rudenko}}, \bibinfo {author}
  {\bibfnamefont {C.}~\bibnamefont {Nicolas}}, \bibinfo {author} {\bibfnamefont
  {X.-J.}\ \bibnamefont {Liu}}, \bibinfo {author} {\bibfnamefont
  {C.}~\bibnamefont {Miron}}, \bibinfo {author} {\bibfnamefont
  {Y.}~\bibnamefont {Zhang}}, \bibinfo {author} {\bibfnamefont
  {Y.}~\bibnamefont {Jiang}}, \bibinfo {author} {\bibfnamefont
  {J.}~\bibnamefont {Chen}}, \bibinfo {author} {\bibfnamefont {M.}~\bibnamefont
  {Anand}}, \bibinfo {author} {\bibfnamefont {D.~E.}\ \bibnamefont {Kim}},
  \bibinfo {author} {\bibfnamefont {K.}~\bibnamefont {Tono}}, \bibinfo {author}
  {\bibfnamefont {M.}~\bibnamefont {Yabashi}}, \bibinfo {author} {\bibfnamefont
  {M.}~\bibnamefont {Yao}}, \ and\ \bibinfo {author} {\bibfnamefont
  {K.}~\bibnamefont {Ueda}},\ }\href {\doibase 10.1021/acs.jpclett.5b01205}
  {\bibfield  {journal} {\bibinfo  {journal} {J. Phys. Chem. Lett.}\ }\textbf
  {\bibinfo {volume} {6}},\ \bibinfo {pages} {2944} (\bibinfo {year}
  {2015})}\BibitemShut {NoStop}%
\bibitem [{\citenamefont {Erk}\ \emph {et~al.}(2014)\citenamefont {Erk},
  \citenamefont {Boll}, \citenamefont {Trippel}, \citenamefont {Anielski},
  \citenamefont {Foucar}, \citenamefont {Rudek}, \citenamefont {Epp},
  \citenamefont {Coffee}, \citenamefont {Carron}, \citenamefont {Schorb},
  \citenamefont {Ferguson}, \citenamefont {Swiggers}, \citenamefont {Bozek},
  \citenamefont {Simon}, \citenamefont {Marchenko}, \citenamefont {K{\"u}pper},
  \citenamefont {Schlichting}, \citenamefont {Ullrich}, \citenamefont
  {Bostedt}, \citenamefont {Rolles},\ and\ \citenamefont {Rudenko}}]{Erk14}%
  \BibitemOpen
  \bibfield  {author} {\bibinfo {author} {\bibfnamefont {B.}~\bibnamefont
  {Erk}}, \bibinfo {author} {\bibfnamefont {R.}~\bibnamefont {Boll}}, \bibinfo
  {author} {\bibfnamefont {S.}~\bibnamefont {Trippel}}, \bibinfo {author}
  {\bibfnamefont {D.}~\bibnamefont {Anielski}}, \bibinfo {author}
  {\bibfnamefont {L.}~\bibnamefont {Foucar}}, \bibinfo {author} {\bibfnamefont
  {B.}~\bibnamefont {Rudek}}, \bibinfo {author} {\bibfnamefont {S.~W.}\
  \bibnamefont {Epp}}, \bibinfo {author} {\bibfnamefont {R.}~\bibnamefont
  {Coffee}}, \bibinfo {author} {\bibfnamefont {S.}~\bibnamefont {Carron}},
  \bibinfo {author} {\bibfnamefont {S.}~\bibnamefont {Schorb}}, \bibinfo
  {author} {\bibfnamefont {K.~R.}\ \bibnamefont {Ferguson}}, \bibinfo {author}
  {\bibfnamefont {M.}~\bibnamefont {Swiggers}}, \bibinfo {author}
  {\bibfnamefont {J.~D.}\ \bibnamefont {Bozek}}, \bibinfo {author}
  {\bibfnamefont {M.}~\bibnamefont {Simon}}, \bibinfo {author} {\bibfnamefont
  {T.}~\bibnamefont {Marchenko}}, \bibinfo {author} {\bibfnamefont
  {J.}~\bibnamefont {K{\"u}pper}}, \bibinfo {author} {\bibfnamefont
  {I.}~\bibnamefont {Schlichting}}, \bibinfo {author} {\bibfnamefont
  {J.}~\bibnamefont {Ullrich}}, \bibinfo {author} {\bibfnamefont
  {C.}~\bibnamefont {Bostedt}}, \bibinfo {author} {\bibfnamefont
  {D.}~\bibnamefont {Rolles}}, \ and\ \bibinfo {author} {\bibfnamefont
  {A.}~\bibnamefont {Rudenko}},\ }\href {\doibase 10.1126/science.1253607}
  {\bibfield  {journal} {\bibinfo  {journal} {Science}\ }\textbf {\bibinfo
  {volume} {345}},\ \bibinfo {pages} {288} (\bibinfo {year}
  {2014})}\BibitemShut {NoStop}%
\bibitem [{\citenamefont {Rudenko}\ \emph {et~al.}(2017)\citenamefont
  {Rudenko}, \citenamefont {Inhester}, \citenamefont {Hanasaki}, \citenamefont
  {Li}, \citenamefont {Robatjazi}, \citenamefont {Erk}, \citenamefont {Boll},
  \citenamefont {Toyota}, \citenamefont {Hao}, \citenamefont {Vendrell},
  \citenamefont {Bomme}, \citenamefont {Savelyev}, \citenamefont {Rudek},
  \citenamefont {Foucar}, \citenamefont {Southworth}, \citenamefont {Lehmann},
  \citenamefont {Kr{\"a}ssig}, \citenamefont {Marchenko}, \citenamefont
  {Simon}, \citenamefont {Ueda}, \citenamefont {Ferguson}, \citenamefont
  {Bucher}, \citenamefont {Gorkhover}, \citenamefont {Carron}, \citenamefont
  {Alonso-Mori}, \citenamefont {Koglin}, \citenamefont {Correa}, \citenamefont
  {Williams}, \citenamefont {Boutet}, \citenamefont {Young}, \citenamefont
  {Bostedt}, \citenamefont {Son}, \citenamefont {Santra},\ and\ \citenamefont
  {Rolles}}]{Rudenko17}%
  \BibitemOpen
  \bibfield  {author} {\bibinfo {author} {\bibfnamefont {A.}~\bibnamefont
  {Rudenko}}, \bibinfo {author} {\bibfnamefont {L.}~\bibnamefont {Inhester}},
  \bibinfo {author} {\bibfnamefont {K.}~\bibnamefont {Hanasaki}}, \bibinfo
  {author} {\bibfnamefont {X.}~\bibnamefont {Li}}, \bibinfo {author}
  {\bibfnamefont {S.~J.}\ \bibnamefont {Robatjazi}}, \bibinfo {author}
  {\bibfnamefont {B.}~\bibnamefont {Erk}}, \bibinfo {author} {\bibfnamefont
  {R.}~\bibnamefont {Boll}}, \bibinfo {author} {\bibfnamefont {K.}~\bibnamefont
  {Toyota}}, \bibinfo {author} {\bibfnamefont {Y.}~\bibnamefont {Hao}},
  \bibinfo {author} {\bibfnamefont {O.}~\bibnamefont {Vendrell}}, \bibinfo
  {author} {\bibfnamefont {C.}~\bibnamefont {Bomme}}, \bibinfo {author}
  {\bibfnamefont {E.}~\bibnamefont {Savelyev}}, \bibinfo {author}
  {\bibfnamefont {B.}~\bibnamefont {Rudek}}, \bibinfo {author} {\bibfnamefont
  {L.}~\bibnamefont {Foucar}}, \bibinfo {author} {\bibfnamefont {S.~H.}\
  \bibnamefont {Southworth}}, \bibinfo {author} {\bibfnamefont {C.~S.}\
  \bibnamefont {Lehmann}}, \bibinfo {author} {\bibfnamefont {B.}~\bibnamefont
  {Kr{\"a}ssig}}, \bibinfo {author} {\bibfnamefont {T.}~\bibnamefont
  {Marchenko}}, \bibinfo {author} {\bibfnamefont {M.}~\bibnamefont {Simon}},
  \bibinfo {author} {\bibfnamefont {K.}~\bibnamefont {Ueda}}, \bibinfo {author}
  {\bibfnamefont {K.~R.}\ \bibnamefont {Ferguson}}, \bibinfo {author}
  {\bibfnamefont {M.}~\bibnamefont {Bucher}}, \bibinfo {author} {\bibfnamefont
  {T.}~\bibnamefont {Gorkhover}}, \bibinfo {author} {\bibfnamefont
  {S.}~\bibnamefont {Carron}}, \bibinfo {author} {\bibfnamefont
  {R.}~\bibnamefont {Alonso-Mori}}, \bibinfo {author} {\bibfnamefont {J.~E.}\
  \bibnamefont {Koglin}}, \bibinfo {author} {\bibfnamefont {J.}~\bibnamefont
  {Correa}}, \bibinfo {author} {\bibfnamefont {G.~J.}\ \bibnamefont
  {Williams}}, \bibinfo {author} {\bibfnamefont {S.}~\bibnamefont {Boutet}},
  \bibinfo {author} {\bibfnamefont {L.}~\bibnamefont {Young}}, \bibinfo
  {author} {\bibfnamefont {C.}~\bibnamefont {Bostedt}}, \bibinfo {author}
  {\bibfnamefont {S.-K.}\ \bibnamefont {Son}}, \bibinfo {author} {\bibfnamefont
  {R.}~\bibnamefont {Santra}}, \ and\ \bibinfo {author} {\bibfnamefont
  {D.}~\bibnamefont {Rolles}},\ }\href {\doibase 10.1038/nature22373}
  {\bibfield  {journal} {\bibinfo  {journal} {Nature}\ }\textbf {\bibinfo
  {volume} {546}},\ \bibinfo {pages} {129} (\bibinfo {year}
  {2017})}\BibitemShut {NoStop}%
\bibitem [{\citenamefont {Murphy}\ \emph {et~al.}(2014)\citenamefont {Murphy},
  \citenamefont {Osipov}, \citenamefont {Jurek}, \citenamefont {Fang},
  \citenamefont {Son}, \citenamefont {Mucke}, \citenamefont {Eland},
  \citenamefont {Zhaunerchyk}, \citenamefont {Feifel}, \citenamefont {Avaldi},
  \citenamefont {Bolognesi}, \citenamefont {Bostedt}, \citenamefont {Bozek},
  \citenamefont {Grilj}, \citenamefont {Guehr}, \citenamefont {Frasinski},
  \citenamefont {Glownia}, \citenamefont {Ha}, \citenamefont {Hoffmann},
  \citenamefont {Kukk}, \citenamefont {McFarland}, \citenamefont {Miron},
  \citenamefont {Sistrunk}, \citenamefont {Squibb}, \citenamefont {Ueda},
  \citenamefont {Santra},\ and\ \citenamefont {Berrah}}]{Murphy14}%
  \BibitemOpen
  \bibfield  {author} {\bibinfo {author} {\bibfnamefont {B.}~\bibnamefont
  {Murphy}}, \bibinfo {author} {\bibfnamefont {T.}~\bibnamefont {Osipov}},
  \bibinfo {author} {\bibfnamefont {Z.}~\bibnamefont {Jurek}}, \bibinfo
  {author} {\bibfnamefont {L.}~\bibnamefont {Fang}}, \bibinfo {author}
  {\bibfnamefont {S.-K.}\ \bibnamefont {Son}}, \bibinfo {author} {\bibfnamefont
  {M.}~\bibnamefont {Mucke}}, \bibinfo {author} {\bibfnamefont {J.~H.~D.}\
  \bibnamefont {Eland}}, \bibinfo {author} {\bibfnamefont {V.}~\bibnamefont
  {Zhaunerchyk}}, \bibinfo {author} {\bibfnamefont {R.}~\bibnamefont {Feifel}},
  \bibinfo {author} {\bibfnamefont {L.}~\bibnamefont {Avaldi}}, \bibinfo
  {author} {\bibfnamefont {P.}~\bibnamefont {Bolognesi}}, \bibinfo {author}
  {\bibfnamefont {C.}~\bibnamefont {Bostedt}}, \bibinfo {author} {\bibfnamefont
  {J.~D.}\ \bibnamefont {Bozek}}, \bibinfo {author} {\bibfnamefont
  {J.}~\bibnamefont {Grilj}}, \bibinfo {author} {\bibfnamefont
  {M.}~\bibnamefont {Guehr}}, \bibinfo {author} {\bibfnamefont {L.~J.}\
  \bibnamefont {Frasinski}}, \bibinfo {author} {\bibfnamefont {J.}~\bibnamefont
  {Glownia}}, \bibinfo {author} {\bibfnamefont {D.~T.}\ \bibnamefont {Ha}},
  \bibinfo {author} {\bibfnamefont {K.}~\bibnamefont {Hoffmann}}, \bibinfo
  {author} {\bibfnamefont {E.}~\bibnamefont {Kukk}}, \bibinfo {author}
  {\bibfnamefont {B.~K.}\ \bibnamefont {McFarland}}, \bibinfo {author}
  {\bibfnamefont {C.}~\bibnamefont {Miron}}, \bibinfo {author} {\bibfnamefont
  {E.}~\bibnamefont {Sistrunk}}, \bibinfo {author} {\bibfnamefont {R.~J.}\
  \bibnamefont {Squibb}}, \bibinfo {author} {\bibfnamefont {K.}~\bibnamefont
  {Ueda}}, \bibinfo {author} {\bibfnamefont {R.}~\bibnamefont {Santra}}, \ and\
  \bibinfo {author} {\bibfnamefont {N.}~\bibnamefont {Berrah}},\ }\href
  {\doibase 10.1038/ncomms5281} {\bibfield  {journal} {\bibinfo  {journal}
  {Nat. Commun.}\ }\textbf {\bibinfo {volume} {5}},\ \bibinfo {pages} {4281}
  (\bibinfo {year} {2014})}\BibitemShut {NoStop}%
\bibitem [{\citenamefont {Tachibana}\ \emph {et~al.}(2015)\citenamefont
  {Tachibana}, \citenamefont {Jurek}, \citenamefont {Fukuzawa}, \citenamefont
  {Motomura}, \citenamefont {Nagaya}, \citenamefont {Wada}, \citenamefont
  {Johnsson}, \citenamefont {Siano}, \citenamefont {Mondal}, \citenamefont
  {Ito}, \citenamefont {Kimura}, \citenamefont {Sakai}, \citenamefont
  {Matsunami}, \citenamefont {Hayashita}, \citenamefont {Kajikawa},
  \citenamefont {Liu}, \citenamefont {Robert}, \citenamefont {Miron},
  \citenamefont {Feifel}, \citenamefont {Marangos}, \citenamefont {Tono},
  \citenamefont {Inubushi}, \citenamefont {Yabashi}, \citenamefont {Son},
  \citenamefont {Ziaja}, \citenamefont {Yao}, \citenamefont {Santra},\ and\
  \citenamefont {Ueda}}]{Tachibana15}%
  \BibitemOpen
  \bibfield  {author} {\bibinfo {author} {\bibfnamefont {T.}~\bibnamefont
  {Tachibana}}, \bibinfo {author} {\bibfnamefont {Z.}~\bibnamefont {Jurek}},
  \bibinfo {author} {\bibfnamefont {H.}~\bibnamefont {Fukuzawa}}, \bibinfo
  {author} {\bibfnamefont {K.}~\bibnamefont {Motomura}}, \bibinfo {author}
  {\bibfnamefont {K.}~\bibnamefont {Nagaya}}, \bibinfo {author} {\bibfnamefont
  {S.}~\bibnamefont {Wada}}, \bibinfo {author} {\bibfnamefont
  {P.}~\bibnamefont {Johnsson}}, \bibinfo {author} {\bibfnamefont
  {M.}~\bibnamefont {Siano}}, \bibinfo {author} {\bibfnamefont
  {S.}~\bibnamefont {Mondal}}, \bibinfo {author} {\bibfnamefont
  {Y.}~\bibnamefont {Ito}}, \bibinfo {author} {\bibfnamefont {M.}~\bibnamefont
  {Kimura}}, \bibinfo {author} {\bibfnamefont {T.}~\bibnamefont {Sakai}},
  \bibinfo {author} {\bibfnamefont {K.}~\bibnamefont {Matsunami}}, \bibinfo
  {author} {\bibfnamefont {H.}~\bibnamefont {Hayashita}}, \bibinfo {author}
  {\bibfnamefont {J.}~\bibnamefont {Kajikawa}}, \bibinfo {author}
  {\bibfnamefont {X.-J.}\ \bibnamefont {Liu}}, \bibinfo {author} {\bibfnamefont
  {E.}~\bibnamefont {Robert}}, \bibinfo {author} {\bibfnamefont
  {C.}~\bibnamefont {Miron}}, \bibinfo {author} {\bibfnamefont
  {R.}~\bibnamefont {Feifel}}, \bibinfo {author} {\bibfnamefont {J.~P.}\
  \bibnamefont {Marangos}}, \bibinfo {author} {\bibfnamefont {K.}~\bibnamefont
  {Tono}}, \bibinfo {author} {\bibfnamefont {Y.}~\bibnamefont {Inubushi}},
  \bibinfo {author} {\bibfnamefont {M.}~\bibnamefont {Yabashi}}, \bibinfo
  {author} {\bibfnamefont {S.-K.}\ \bibnamefont {Son}}, \bibinfo {author}
  {\bibfnamefont {B.}~\bibnamefont {Ziaja}}, \bibinfo {author} {\bibfnamefont
  {M.}~\bibnamefont {Yao}}, \bibinfo {author} {\bibfnamefont {R.}~\bibnamefont
  {Santra}}, \ and\ \bibinfo {author} {\bibfnamefont {K.}~\bibnamefont
  {Ueda}},\ }\href {\doibase 10.1038/srep10977} {\bibfield  {journal} {\bibinfo
   {journal} {Sci. Rep.}\ }\textbf {\bibinfo {volume} {5}},\ \bibinfo {pages}
  {10977} (\bibinfo {year} {2015})}\BibitemShut {NoStop}%
\bibitem [{\citenamefont {Kumagai}\ \emph {et~al.}(2018)\citenamefont
  {Kumagai}, \citenamefont {Jurek}, \citenamefont {Xu}, \citenamefont
  {Fukuzawa}, \citenamefont {Motomura}, \citenamefont {Iablonskyi},
  \citenamefont {Nagaya}, \citenamefont {Wada}, \citenamefont {Mondal},
  \citenamefont {Tachibana}, \citenamefont {Ito}, \citenamefont {Sakai},
  \citenamefont {Matsunami}, \citenamefont {Nishiyama}, \citenamefont
  {Umemoto}, \citenamefont {Nicolas}, \citenamefont {Miron}, \citenamefont
  {Togashi}, \citenamefont {Ogawa}, \citenamefont {Owada}, \citenamefont
  {Tono}, \citenamefont {Yabashi}, \citenamefont {Son}, \citenamefont {Ziaja},
  \citenamefont {Santra},\ and\ \citenamefont {Ueda}}]{Kumagai18}%
  \BibitemOpen
  \bibfield  {author} {\bibinfo {author} {\bibfnamefont {Y.}~\bibnamefont
  {Kumagai}}, \bibinfo {author} {\bibfnamefont {Z.}~\bibnamefont {Jurek}},
  \bibinfo {author} {\bibfnamefont {W.}~\bibnamefont {Xu}}, \bibinfo {author}
  {\bibfnamefont {H.}~\bibnamefont {Fukuzawa}}, \bibinfo {author}
  {\bibfnamefont {K.}~\bibnamefont {Motomura}}, \bibinfo {author}
  {\bibfnamefont {D.}~\bibnamefont {Iablonskyi}}, \bibinfo {author}
  {\bibfnamefont {K.}~\bibnamefont {Nagaya}}, \bibinfo {author} {\bibfnamefont
  {S.}~\bibnamefont {Wada}}, \bibinfo {author} {\bibfnamefont
  {S.}~\bibnamefont {Mondal}}, \bibinfo {author} {\bibfnamefont
  {T.}~\bibnamefont {Tachibana}}, \bibinfo {author} {\bibfnamefont
  {Y.}~\bibnamefont {Ito}}, \bibinfo {author} {\bibfnamefont {T.}~\bibnamefont
  {Sakai}}, \bibinfo {author} {\bibfnamefont {K.}~\bibnamefont {Matsunami}},
  \bibinfo {author} {\bibfnamefont {T.}~\bibnamefont {Nishiyama}}, \bibinfo
  {author} {\bibfnamefont {T.}~\bibnamefont {Umemoto}}, \bibinfo {author}
  {\bibfnamefont {C.}~\bibnamefont {Nicolas}}, \bibinfo {author} {\bibfnamefont
  {C.}~\bibnamefont {Miron}}, \bibinfo {author} {\bibfnamefont
  {T.}~\bibnamefont {Togashi}}, \bibinfo {author} {\bibfnamefont
  {K.}~\bibnamefont {Ogawa}}, \bibinfo {author} {\bibfnamefont
  {S.}~\bibnamefont {Owada}}, \bibinfo {author} {\bibfnamefont
  {K.}~\bibnamefont {Tono}}, \bibinfo {author} {\bibfnamefont {M.}~\bibnamefont
  {Yabashi}}, \bibinfo {author} {\bibfnamefont {S.-K.}\ \bibnamefont {Son}},
  \bibinfo {author} {\bibfnamefont {B.}~\bibnamefont {Ziaja}}, \bibinfo
  {author} {\bibfnamefont {R.}~\bibnamefont {Santra}}, \ and\ \bibinfo {author}
  {\bibfnamefont {K.}~\bibnamefont {Ueda}},\ }\href {\doibase
  10.1103/PhysRevLett.120.223201} {\bibfield  {journal} {\bibinfo  {journal}
  {Phys. Rev. Lett.}\ }\textbf {\bibinfo {volume} {120}},\ \bibinfo {pages}
  {223201} (\bibinfo {year} {2018})}\BibitemShut {NoStop}%
\bibitem [{\citenamefont {Inhester}\ \emph {et~al.}(2016)\citenamefont
  {Inhester}, \citenamefont {Hanasaki}, \citenamefont {Hao}, \citenamefont
  {Son},\ and\ \citenamefont {Santra}}]{Inhester16}%
  \BibitemOpen
  \bibfield  {author} {\bibinfo {author} {\bibfnamefont {L.}~\bibnamefont
  {Inhester}}, \bibinfo {author} {\bibfnamefont {K.}~\bibnamefont {Hanasaki}},
  \bibinfo {author} {\bibfnamefont {Y.}~\bibnamefont {Hao}}, \bibinfo {author}
  {\bibfnamefont {S.-K.}\ \bibnamefont {Son}}, \ and\ \bibinfo {author}
  {\bibfnamefont {R.}~\bibnamefont {Santra}},\ }\href {\doibase
  10.1103/PhysRevA.94.023422} {\bibfield  {journal} {\bibinfo  {journal} {Phys.
  Rev. A}\ }\textbf {\bibinfo {volume} {94}},\ \bibinfo {pages} {023422}
  (\bibinfo {year} {2016})}\BibitemShut {NoStop}%
\bibitem [{\citenamefont {Choi}(2017)}]{blackhole}%
  \BibitemOpen
  \bibfield  {author} {\bibinfo {author} {\bibfnamefont {C.~Q.}\ \bibnamefont
  {Choi}},\ }\href
  {https://www.scientificamerican.com/article/x-ray-lasers-make-atoms-act-like-ldquo-black-holes-rdquo-in-molecules/}
  {\enquote {\bibinfo {title} {X-ray lasers make atoms act like `black holes'
  in molecules},}\ } {\bibinfo  {journal} {Sci. Am.}\ }(\bibinfo {year} {June 1, 2017})\BibitemShut {NoStop}%
\bibitem [{\citenamefont {Chapman}\ \emph {et~al.}(2011)\citenamefont
  {Chapman}, \citenamefont {Fromme}, \citenamefont {Barty}, \citenamefont
  {White}, \citenamefont {Kirian}, \citenamefont {Aquila}, \citenamefont
  {Hunter}, \citenamefont {Schulz}, \citenamefont {DePonte}, \citenamefont
  {Weierstall}, \citenamefont {Doak}, \citenamefont {Maia}, \citenamefont
  {Martin}, \citenamefont {Schlichting}, \citenamefont {Lomb}, \citenamefont
  {Coppola}, \citenamefont {Shoeman}, \citenamefont {Epp}, \citenamefont
  {Hartmann}, \citenamefont {Rolles}, \citenamefont {Rudenko}, \citenamefont
  {Foucar}, \citenamefont {Kimmel}, \citenamefont {Weidenspointner},
  \citenamefont {Holl}, \citenamefont {Liang}, \citenamefont {Barthelmess},
  \citenamefont {Caleman}, \citenamefont {Boutet}, \citenamefont {Bogan},
  \citenamefont {Krzywinski}, \citenamefont {Bostedt}, \citenamefont {Bajt},
  \citenamefont {Gumprecht}, \citenamefont {Rudek}, \citenamefont {Erk},
  \citenamefont {Schmidt}, \citenamefont {H{\"o}mke}, \citenamefont {Reich},
  \citenamefont {Pietschner}, \citenamefont {Str{\"u}der}, \citenamefont
  {Hauser}, \citenamefont {Gorke}, \citenamefont {Ullrich}, \citenamefont
  {Herrmann}, \citenamefont {Schaller}, \citenamefont {Schopper}, \citenamefont
  {Soltau}, \citenamefont {K{\"u}hnel}, \citenamefont {Messerschmidt},
  \citenamefont {Bozek}, \citenamefont {Hau-Riege}, \citenamefont {Frank},
  \citenamefont {Hampton}, \citenamefont {Sierra}, \citenamefont {Starodub},
  \citenamefont {Williams}, \citenamefont {Hajdu}, \citenamefont {Timneanu},
  \citenamefont {Seibert}, \citenamefont {Andreasson}, \citenamefont {Rocker},
  \citenamefont {J{\"o}nsson}, \citenamefont {Svenda}, \citenamefont {Stern},
  \citenamefont {Nass}, \citenamefont {Andritschke}, \citenamefont
  {Schr{\"o}ter}, \citenamefont {Krasniqi}, \citenamefont {Bott}, \citenamefont
  {Schmidt}, \citenamefont {Wang}, \citenamefont {Grotjohann}, \citenamefont
  {Holton}, \citenamefont {Barends}, \citenamefont {Neutze}, \citenamefont
  {Marchesini}, \citenamefont {Fromme}, \citenamefont {Schorb}, \citenamefont
  {Rupp}, \citenamefont {Adolph}, \citenamefont {Gorkhover}, \citenamefont
  {Andersson}, \citenamefont {Hirsemann}, \citenamefont {Potdevin},
  \citenamefont {Graafsma}, \citenamefont {Nilsson},\ and\ \citenamefont
  {Spence}}]{Chapman11}%
  \BibitemOpen
  \bibfield  {author} {\bibinfo {author} {\bibfnamefont {H.~N.}\ \bibnamefont
  {Chapman}}, \bibinfo {author} {\bibfnamefont {P.}~\bibnamefont {Fromme}},
  \bibinfo {author} {\bibfnamefont {A.}~\bibnamefont {Barty}}, \bibinfo
  {author} {\bibfnamefont {T.~A.}\ \bibnamefont {White}}, \bibinfo {author}
  {\bibfnamefont {R.~A.}\ \bibnamefont {Kirian}}, \bibinfo {author}
  {\bibfnamefont {A.}~\bibnamefont {Aquila}}, \bibinfo {author} {\bibfnamefont
  {M.~S.}\ \bibnamefont {Hunter}}, \bibinfo {author} {\bibfnamefont
  {J.}~\bibnamefont {Schulz}}, \bibinfo {author} {\bibfnamefont {D.~P.}\
  \bibnamefont {DePonte}}, \bibinfo {author} {\bibfnamefont {U.}~\bibnamefont
  {Weierstall}}, \bibinfo {author} {\bibfnamefont {R.~B.}\ \bibnamefont
  {Doak}}, \bibinfo {author} {\bibfnamefont {F.~R. N.~C.}\ \bibnamefont
  {Maia}}, \bibinfo {author} {\bibfnamefont {A.~V.}\ \bibnamefont {Martin}},
  \bibinfo {author} {\bibfnamefont {I.}~\bibnamefont {Schlichting}}, \bibinfo
  {author} {\bibfnamefont {L.}~\bibnamefont {Lomb}}, \bibinfo {author}
  {\bibfnamefont {N.}~\bibnamefont {Coppola}}, \bibinfo {author} {\bibfnamefont
  {R.~L.}\ \bibnamefont {Shoeman}}, \bibinfo {author} {\bibfnamefont {S.~W.}\
  \bibnamefont {Epp}}, \bibinfo {author} {\bibfnamefont {R.}~\bibnamefont
  {Hartmann}}, \bibinfo {author} {\bibfnamefont {D.}~\bibnamefont {Rolles}},
  \bibinfo {author} {\bibfnamefont {A.}~\bibnamefont {Rudenko}}, \bibinfo
  {author} {\bibfnamefont {L.}~\bibnamefont {Foucar}}, \bibinfo {author}
  {\bibfnamefont {N.}~\bibnamefont {Kimmel}}, \bibinfo {author} {\bibfnamefont
  {G.}~\bibnamefont {Weidenspointner}}, \bibinfo {author} {\bibfnamefont
  {P.}~\bibnamefont {Holl}}, \bibinfo {author} {\bibfnamefont {M.}~\bibnamefont
  {Liang}}, \bibinfo {author} {\bibfnamefont {M.}~\bibnamefont {Barthelmess}},
  \bibinfo {author} {\bibfnamefont {C.}~\bibnamefont {Caleman}}, \bibinfo
  {author} {\bibfnamefont {S.}~\bibnamefont {Boutet}}, \bibinfo {author}
  {\bibfnamefont {M.~J.}\ \bibnamefont {Bogan}}, \bibinfo {author}
  {\bibfnamefont {J.}~\bibnamefont {Krzywinski}}, \bibinfo {author}
  {\bibfnamefont {C.}~\bibnamefont {Bostedt}}, \bibinfo {author} {\bibfnamefont
  {S.}~\bibnamefont {Bajt}}, \bibinfo {author} {\bibfnamefont {L.}~\bibnamefont
  {Gumprecht}}, \bibinfo {author} {\bibfnamefont {B.}~\bibnamefont {Rudek}},
  \bibinfo {author} {\bibfnamefont {B.}~\bibnamefont {Erk}}, \bibinfo {author}
  {\bibfnamefont {C.}~\bibnamefont {Schmidt}}, \bibinfo {author} {\bibfnamefont
  {A.}~\bibnamefont {H{\"o}mke}}, \bibinfo {author} {\bibfnamefont
  {C.}~\bibnamefont {Reich}}, \bibinfo {author} {\bibfnamefont
  {D.}~\bibnamefont {Pietschner}}, \bibinfo {author} {\bibfnamefont
  {L.}~\bibnamefont {Str{\"u}der}}, \bibinfo {author} {\bibfnamefont
  {G.}~\bibnamefont {Hauser}}, \bibinfo {author} {\bibfnamefont
  {H.}~\bibnamefont {Gorke}}, \bibinfo {author} {\bibfnamefont
  {J.}~\bibnamefont {Ullrich}}, \bibinfo {author} {\bibfnamefont
  {S.}~\bibnamefont {Herrmann}}, \bibinfo {author} {\bibfnamefont
  {G.}~\bibnamefont {Schaller}}, \bibinfo {author} {\bibfnamefont
  {F.}~\bibnamefont {Schopper}}, \bibinfo {author} {\bibfnamefont
  {H.}~\bibnamefont {Soltau}}, \bibinfo {author} {\bibfnamefont {K.-U.}\
  \bibnamefont {K{\"u}hnel}}, \bibinfo {author} {\bibfnamefont
  {M.}~\bibnamefont {Messerschmidt}}, \bibinfo {author} {\bibfnamefont {J.~D.}\
  \bibnamefont {Bozek}}, \bibinfo {author} {\bibfnamefont {S.~P.}\ \bibnamefont
  {Hau-Riege}}, \bibinfo {author} {\bibfnamefont {M.}~\bibnamefont {Frank}},
  \bibinfo {author} {\bibfnamefont {C.~Y.}\ \bibnamefont {Hampton}}, \bibinfo
  {author} {\bibfnamefont {R.~G.}\ \bibnamefont {Sierra}}, \bibinfo {author}
  {\bibfnamefont {D.}~\bibnamefont {Starodub}}, \bibinfo {author}
  {\bibfnamefont {G.~J.}\ \bibnamefont {Williams}}, \bibinfo {author}
  {\bibfnamefont {J.}~\bibnamefont {Hajdu}}, \bibinfo {author} {\bibfnamefont
  {N.}~\bibnamefont {Timneanu}}, \bibinfo {author} {\bibfnamefont {M.~M.}\
  \bibnamefont {Seibert}}, \bibinfo {author} {\bibfnamefont {J.}~\bibnamefont
  {Andreasson}}, \bibinfo {author} {\bibfnamefont {A.}~\bibnamefont {Rocker}},
  \bibinfo {author} {\bibfnamefont {O.}~\bibnamefont {J{\"o}nsson}}, \bibinfo
  {author} {\bibfnamefont {M.}~\bibnamefont {Svenda}}, \bibinfo {author}
  {\bibfnamefont {S.}~\bibnamefont {Stern}}, \bibinfo {author} {\bibfnamefont
  {K.}~\bibnamefont {Nass}}, \bibinfo {author} {\bibfnamefont {R.}~\bibnamefont
  {Andritschke}}, \bibinfo {author} {\bibfnamefont {C.-D.}\ \bibnamefont
  {Schr{\"o}ter}}, \bibinfo {author} {\bibfnamefont {F.}~\bibnamefont
  {Krasniqi}}, \bibinfo {author} {\bibfnamefont {M.}~\bibnamefont {Bott}},
  \bibinfo {author} {\bibfnamefont {K.~E.}\ \bibnamefont {Schmidt}}, \bibinfo
  {author} {\bibfnamefont {X.}~\bibnamefont {Wang}}, \bibinfo {author}
  {\bibfnamefont {I.}~\bibnamefont {Grotjohann}}, \bibinfo {author}
  {\bibfnamefont {J.~M.}\ \bibnamefont {Holton}}, \bibinfo {author}
  {\bibfnamefont {T.~R.~M.}\ \bibnamefont {Barends}}, \bibinfo {author}
  {\bibfnamefont {R.}~\bibnamefont {Neutze}}, \bibinfo {author} {\bibfnamefont
  {S.}~\bibnamefont {Marchesini}}, \bibinfo {author} {\bibfnamefont
  {R.}~\bibnamefont {Fromme}}, \bibinfo {author} {\bibfnamefont
  {S.}~\bibnamefont {Schorb}}, \bibinfo {author} {\bibfnamefont
  {D.}~\bibnamefont {Rupp}}, \bibinfo {author} {\bibfnamefont {M.}~\bibnamefont
  {Adolph}}, \bibinfo {author} {\bibfnamefont {T.}~\bibnamefont {Gorkhover}},
  \bibinfo {author} {\bibfnamefont {I.}~\bibnamefont {Andersson}}, \bibinfo
  {author} {\bibfnamefont {H.}~\bibnamefont {Hirsemann}}, \bibinfo {author}
  {\bibfnamefont {G.}~\bibnamefont {Potdevin}}, \bibinfo {author}
  {\bibfnamefont {H.}~\bibnamefont {Graafsma}}, \bibinfo {author}
  {\bibfnamefont {B.}~\bibnamefont {Nilsson}}, \ and\ \bibinfo {author}
  {\bibfnamefont {J.~C.~H.}\ \bibnamefont {Spence}},\ }\href@noop {} {\bibfield
   {journal} {\bibinfo  {journal} {Nature}\ }\textbf {\bibinfo {volume}
  {470}},\ \bibinfo {pages} {73} (\bibinfo {year} {2011})}\BibitemShut
  {NoStop}%
\bibitem [{\citenamefont {Boutet}\ \emph {et~al.}(2012)\citenamefont {Boutet},
  \citenamefont {Lomb}, \citenamefont {Williams}, \citenamefont {Barends},
  \citenamefont {Aquila}, \citenamefont {Doak}, \citenamefont {Weierstall},
  \citenamefont {DePonte}, \citenamefont {Steinbrener}, \citenamefont
  {Shoeman}, \citenamefont {Messerschmidt}, \citenamefont {Barty},
  \citenamefont {White}, \citenamefont {Kassemeyer}, \citenamefont {Kirian},
  \citenamefont {Seibert}, \citenamefont {Montanez}, \citenamefont {Kenney},
  \citenamefont {Herbst}, \citenamefont {Hart}, \citenamefont {Pines},
  \citenamefont {Haller}, \citenamefont {Gruner}, \citenamefont {Philipp},
  \citenamefont {Tate}, \citenamefont {Hromalik}, \citenamefont {Koerner},
  \citenamefont {van Bakel}, \citenamefont {Morse}, \citenamefont {Ghonsalves},
  \citenamefont {Arnlund}, \citenamefont {Bogan}, \citenamefont {Caleman},
  \citenamefont {Fromme}, \citenamefont {Hampton}, \citenamefont {Hunter},
  \citenamefont {Johansson}, \citenamefont {Katona}, \citenamefont {Kupitz},
  \citenamefont {Liang}, \citenamefont {Martin}, \citenamefont {Nass},
  \citenamefont {Redecke}, \citenamefont {Stellato}, \citenamefont {Timneanu},
  \citenamefont {Wang}, \citenamefont {Zatsepin}, \citenamefont {Schafer},
  \citenamefont {Defever}, \citenamefont {Neutze}, \citenamefont {Fromme},
  \citenamefont {Spence}, \citenamefont {Chapman},\ and\ \citenamefont
  {Schlichting}}]{Boutet12}%
  \BibitemOpen
  \bibfield  {author} {\bibinfo {author} {\bibfnamefont {S.}~\bibnamefont
  {Boutet}}, \bibinfo {author} {\bibfnamefont {L.}~\bibnamefont {Lomb}},
  \bibinfo {author} {\bibfnamefont {G.~J.}\ \bibnamefont {Williams}}, \bibinfo
  {author} {\bibfnamefont {T.~R.~M.}\ \bibnamefont {Barends}}, \bibinfo
  {author} {\bibfnamefont {A.}~\bibnamefont {Aquila}}, \bibinfo {author}
  {\bibfnamefont {R.~B.}\ \bibnamefont {Doak}}, \bibinfo {author}
  {\bibfnamefont {U.}~\bibnamefont {Weierstall}}, \bibinfo {author}
  {\bibfnamefont {D.~P.}\ \bibnamefont {DePonte}}, \bibinfo {author}
  {\bibfnamefont {J.}~\bibnamefont {Steinbrener}}, \bibinfo {author}
  {\bibfnamefont {R.~L.}\ \bibnamefont {Shoeman}}, \bibinfo {author}
  {\bibfnamefont {M.}~\bibnamefont {Messerschmidt}}, \bibinfo {author}
  {\bibfnamefont {A.}~\bibnamefont {Barty}}, \bibinfo {author} {\bibfnamefont
  {T.~A.}\ \bibnamefont {White}}, \bibinfo {author} {\bibfnamefont
  {S.}~\bibnamefont {Kassemeyer}}, \bibinfo {author} {\bibfnamefont {R.~A.}\
  \bibnamefont {Kirian}}, \bibinfo {author} {\bibfnamefont {M.~M.}\
  \bibnamefont {Seibert}}, \bibinfo {author} {\bibfnamefont {P.~A.}\
  \bibnamefont {Montanez}}, \bibinfo {author} {\bibfnamefont {C.}~\bibnamefont
  {Kenney}}, \bibinfo {author} {\bibfnamefont {R.}~\bibnamefont {Herbst}},
  \bibinfo {author} {\bibfnamefont {P.}~\bibnamefont {Hart}}, \bibinfo {author}
  {\bibfnamefont {J.}~\bibnamefont {Pines}}, \bibinfo {author} {\bibfnamefont
  {G.}~\bibnamefont {Haller}}, \bibinfo {author} {\bibfnamefont {S.~M.}\
  \bibnamefont {Gruner}}, \bibinfo {author} {\bibfnamefont {H.~T.}\
  \bibnamefont {Philipp}}, \bibinfo {author} {\bibfnamefont {M.~W.}\
  \bibnamefont {Tate}}, \bibinfo {author} {\bibfnamefont {M.}~\bibnamefont
  {Hromalik}}, \bibinfo {author} {\bibfnamefont {L.~J.}\ \bibnamefont
  {Koerner}}, \bibinfo {author} {\bibfnamefont {N.}~\bibnamefont {van Bakel}},
  \bibinfo {author} {\bibfnamefont {J.}~\bibnamefont {Morse}}, \bibinfo
  {author} {\bibfnamefont {W.}~\bibnamefont {Ghonsalves}}, \bibinfo {author}
  {\bibfnamefont {D.}~\bibnamefont {Arnlund}}, \bibinfo {author} {\bibfnamefont
  {M.~J.}\ \bibnamefont {Bogan}}, \bibinfo {author} {\bibfnamefont
  {C.}~\bibnamefont {Caleman}}, \bibinfo {author} {\bibfnamefont
  {R.}~\bibnamefont {Fromme}}, \bibinfo {author} {\bibfnamefont {C.~Y.}\
  \bibnamefont {Hampton}}, \bibinfo {author} {\bibfnamefont {M.~S.}\
  \bibnamefont {Hunter}}, \bibinfo {author} {\bibfnamefont {L.~C.}\
  \bibnamefont {Johansson}}, \bibinfo {author} {\bibfnamefont {G.}~\bibnamefont
  {Katona}}, \bibinfo {author} {\bibfnamefont {C.}~\bibnamefont {Kupitz}},
  \bibinfo {author} {\bibfnamefont {M.}~\bibnamefont {Liang}}, \bibinfo
  {author} {\bibfnamefont {A.~V.}\ \bibnamefont {Martin}}, \bibinfo {author}
  {\bibfnamefont {K.}~\bibnamefont {Nass}}, \bibinfo {author} {\bibfnamefont
  {L.}~\bibnamefont {Redecke}}, \bibinfo {author} {\bibfnamefont
  {F.}~\bibnamefont {Stellato}}, \bibinfo {author} {\bibfnamefont
  {N.}~\bibnamefont {Timneanu}}, \bibinfo {author} {\bibfnamefont
  {D.}~\bibnamefont {Wang}}, \bibinfo {author} {\bibfnamefont {N.~A.}\
  \bibnamefont {Zatsepin}}, \bibinfo {author} {\bibfnamefont {D.}~\bibnamefont
  {Schafer}}, \bibinfo {author} {\bibfnamefont {J.}~\bibnamefont {Defever}},
  \bibinfo {author} {\bibfnamefont {R.}~\bibnamefont {Neutze}}, \bibinfo
  {author} {\bibfnamefont {P.}~\bibnamefont {Fromme}}, \bibinfo {author}
  {\bibfnamefont {J.~C.~H.}\ \bibnamefont {Spence}}, \bibinfo {author}
  {\bibfnamefont {H.~N.}\ \bibnamefont {Chapman}}, \ and\ \bibinfo {author}
  {\bibfnamefont {I.}~\bibnamefont {Schlichting}},\ }\href@noop {} {\bibfield
  {journal} {\bibinfo  {journal} {Science}\ }\textbf {\bibinfo {volume}
  {337}},\ \bibinfo {pages} {362} (\bibinfo {year} {2012})}\BibitemShut
  {NoStop}%
\bibitem [{\citenamefont {Aquila}\ \emph {et~al.}(2015)\citenamefont {Aquila},
  \citenamefont {Barty}, \citenamefont {Bostedt}, \citenamefont {Boutet},
  \citenamefont {Carini}, \citenamefont {dePonte}, \citenamefont {Drell},
  \citenamefont {Doniach}, \citenamefont {Downing}, \citenamefont {Earnest},
  \citenamefont {Elmlund}, \citenamefont {Elser}, \citenamefont {G{\"u}hr},
  \citenamefont {Hajdu}, \citenamefont {Hastings}, \citenamefont {Hau-Riege},
  \citenamefont {Huang}, \citenamefont {Lattman}, \citenamefont {Maia},
  \citenamefont {Marchesini}, \citenamefont {Ourmazd}, \citenamefont
  {Pellegrini}, \citenamefont {Santra}, \citenamefont {Schlichting},
  \citenamefont {Schroer}, \citenamefont {Spence}, \citenamefont {Vartanyants},
  \citenamefont {Wakatsuki}, \citenamefont {Weis},\ and\ \citenamefont
  {Williams}}]{Aquila15}%
  \BibitemOpen
  \bibfield  {author} {\bibinfo {author} {\bibfnamefont {A.}~\bibnamefont
  {Aquila}}, \bibinfo {author} {\bibfnamefont {A.}~\bibnamefont {Barty}},
  \bibinfo {author} {\bibfnamefont {C.}~\bibnamefont {Bostedt}}, \bibinfo
  {author} {\bibfnamefont {S.}~\bibnamefont {Boutet}}, \bibinfo {author}
  {\bibfnamefont {G.}~\bibnamefont {Carini}}, \bibinfo {author} {\bibfnamefont
  {D.}~\bibnamefont {dePonte}}, \bibinfo {author} {\bibfnamefont
  {P.}~\bibnamefont {Drell}}, \bibinfo {author} {\bibfnamefont
  {S.}~\bibnamefont {Doniach}}, \bibinfo {author} {\bibfnamefont {K.~H.}\
  \bibnamefont {Downing}}, \bibinfo {author} {\bibfnamefont {T.}~\bibnamefont
  {Earnest}}, \bibinfo {author} {\bibfnamefont {H.}~\bibnamefont {Elmlund}},
  \bibinfo {author} {\bibfnamefont {V.}~\bibnamefont {Elser}}, \bibinfo
  {author} {\bibfnamefont {M.}~\bibnamefont {G{\"u}hr}}, \bibinfo {author}
  {\bibfnamefont {J.}~\bibnamefont {Hajdu}}, \bibinfo {author} {\bibfnamefont
  {J.}~\bibnamefont {Hastings}}, \bibinfo {author} {\bibfnamefont {S.~P.}\
  \bibnamefont {Hau-Riege}}, \bibinfo {author} {\bibfnamefont {Z.}~\bibnamefont
  {Huang}}, \bibinfo {author} {\bibfnamefont {E.~E.}\ \bibnamefont {Lattman}},
  \bibinfo {author} {\bibfnamefont {F.~R. N.~C.}\ \bibnamefont {Maia}},
  \bibinfo {author} {\bibfnamefont {S.}~\bibnamefont {Marchesini}}, \bibinfo
  {author} {\bibfnamefont {A.}~\bibnamefont {Ourmazd}}, \bibinfo {author}
  {\bibfnamefont {C.}~\bibnamefont {Pellegrini}}, \bibinfo {author}
  {\bibfnamefont {R.}~\bibnamefont {Santra}}, \bibinfo {author} {\bibfnamefont
  {I.}~\bibnamefont {Schlichting}}, \bibinfo {author} {\bibfnamefont
  {C.}~\bibnamefont {Schroer}}, \bibinfo {author} {\bibfnamefont {J.~C.~H.}\
  \bibnamefont {Spence}}, \bibinfo {author} {\bibfnamefont {I.~A.}\
  \bibnamefont {Vartanyants}}, \bibinfo {author} {\bibfnamefont
  {S.}~\bibnamefont {Wakatsuki}}, \bibinfo {author} {\bibfnamefont {W.~I.}\
  \bibnamefont {Weis}}, \ and\ \bibinfo {author} {\bibfnamefont {G.~J.}\
  \bibnamefont {Williams}},\ }\href {\doibase 10.1063/1.4918726} {\bibfield
  {journal} {\bibinfo  {journal} {Struct. Dyn.}\ }\textbf {\bibinfo {volume}
  {2}},\ \bibinfo {pages} {041701} (\bibinfo {year} {2015})}\BibitemShut
  {NoStop}%
\bibitem [{\citenamefont {Yoon}\ \emph {et~al.}(2016)\citenamefont {Yoon},
  \citenamefont {Yurkov}, \citenamefont {Schneidmiller}, \citenamefont
  {Samoylova}, \citenamefont {Buzmakov}, \citenamefont {Jurek}, \citenamefont
  {Ziaja}, \citenamefont {Santra}, \citenamefont {Loh}, \citenamefont
  {Tschentscher},\ and\ \citenamefont {Mancuso}}]{Yoon16}%
  \BibitemOpen
  \bibfield  {author} {\bibinfo {author} {\bibfnamefont {C.~H.}\ \bibnamefont
  {Yoon}}, \bibinfo {author} {\bibfnamefont {M.~V.}\ \bibnamefont {Yurkov}},
  \bibinfo {author} {\bibfnamefont {E.~A.}\ \bibnamefont {Schneidmiller}},
  \bibinfo {author} {\bibfnamefont {L.}~\bibnamefont {Samoylova}}, \bibinfo
  {author} {\bibfnamefont {A.}~\bibnamefont {Buzmakov}}, \bibinfo {author}
  {\bibfnamefont {Z.}~\bibnamefont {Jurek}}, \bibinfo {author} {\bibfnamefont
  {B.}~\bibnamefont {Ziaja}}, \bibinfo {author} {\bibfnamefont
  {R.}~\bibnamefont {Santra}}, \bibinfo {author} {\bibfnamefont {N.~D.}\
  \bibnamefont {Loh}}, \bibinfo {author} {\bibfnamefont {T.}~\bibnamefont
  {Tschentscher}}, \ and\ \bibinfo {author} {\bibfnamefont {A.~P.}\
  \bibnamefont {Mancuso}},\ }\href {\doibase 10.1038/srep24791} {\bibfield
  {journal} {\bibinfo  {journal} {Sci. Rep.}\ }\textbf {\bibinfo {volume}
  {6}},\ \bibinfo {pages} {24791} (\bibinfo {year} {2016})}\BibitemShut
  {NoStop}%
\bibitem [{\citenamefont {Fortmann-Grote}\ \emph {et~al.}(2017)\citenamefont
  {Fortmann-Grote}, \citenamefont {Buzmakov}, \citenamefont {Jurek},
  \citenamefont {Loh}, \citenamefont {Samoylova}, \citenamefont {Santra},
  \citenamefont {Schneidmiller}, \citenamefont {Tschentscher}, \citenamefont
  {Yakubov}, \citenamefont {Yoon}, \citenamefont {Yurkov}, \citenamefont
  {Ziaja-Motyka},\ and\ \citenamefont {Mancuso}}]{Fortmann-Grote17}%
  \BibitemOpen
  \bibfield  {author} {\bibinfo {author} {\bibfnamefont {C.}~\bibnamefont
  {Fortmann-Grote}}, \bibinfo {author} {\bibfnamefont {A.}~\bibnamefont
  {Buzmakov}}, \bibinfo {author} {\bibfnamefont {Z.}~\bibnamefont {Jurek}},
  \bibinfo {author} {\bibfnamefont {N.-T.~D.}\ \bibnamefont {Loh}}, \bibinfo
  {author} {\bibfnamefont {L.}~\bibnamefont {Samoylova}}, \bibinfo {author}
  {\bibfnamefont {R.}~\bibnamefont {Santra}}, \bibinfo {author} {\bibfnamefont
  {E.~A.}\ \bibnamefont {Schneidmiller}}, \bibinfo {author} {\bibfnamefont
  {T.}~\bibnamefont {Tschentscher}}, \bibinfo {author} {\bibfnamefont
  {S.}~\bibnamefont {Yakubov}}, \bibinfo {author} {\bibfnamefont {C.~H.}\
  \bibnamefont {Yoon}}, \bibinfo {author} {\bibfnamefont {M.~V.}\ \bibnamefont
  {Yurkov}}, \bibinfo {author} {\bibfnamefont {B.}~\bibnamefont
  {Ziaja-Motyka}}, \ and\ \bibinfo {author} {\bibfnamefont {A.~P.}\
  \bibnamefont {Mancuso}},\ }\href {\doibase 10.1107/S2052252517009496}
  {\bibfield  {journal} {\bibinfo  {journal} {IUCrJ}\ }\textbf {\bibinfo
  {volume} {4}},\ \bibinfo {pages} {560} (\bibinfo {year} {2017})}\BibitemShut
  {NoStop}%
\bibitem [{\citenamefont {Son}\ \emph {et~al.}(2011{\natexlab{a}})\citenamefont
  {Son}, \citenamefont {Chapman},\ and\ \citenamefont {Santra}}]{Son11e}%
  \BibitemOpen
  \bibfield  {author} {\bibinfo {author} {\bibfnamefont {S.-K.}\ \bibnamefont
  {Son}}, \bibinfo {author} {\bibfnamefont {H.~N.}\ \bibnamefont {Chapman}}, \
  and\ \bibinfo {author} {\bibfnamefont {R.}~\bibnamefont {Santra}},\ }\href
  {\doibase 10.1103/PhysRevLett.107.218102} {\bibfield  {journal} {\bibinfo
  {journal} {Phys. Rev. Lett.}\ }\textbf {\bibinfo {volume} {107}},\ \bibinfo
  {pages} {218102} (\bibinfo {year} {2011}{\natexlab{a}})}\BibitemShut
  {NoStop}%
\bibitem [{\citenamefont {Galli}\ \emph {et~al.}(2015)\citenamefont {Galli},
  \citenamefont {Son}, \citenamefont {Barends}, \citenamefont {White},
  \citenamefont {Barty}, \citenamefont {Botha}, \citenamefont {Boutet},
  \citenamefont {Caleman}, \citenamefont {Doak}, \citenamefont {Nanao},
  \citenamefont {Nass}, \citenamefont {Schoeman}, \citenamefont {Timneanu},
  \citenamefont {Santra}, \citenamefont {Schlichting},\ and\ \citenamefont
  {Chapman}}]{Galli15b}%
  \BibitemOpen
  \bibfield  {author} {\bibinfo {author} {\bibfnamefont {L.}~\bibnamefont
  {Galli}}, \bibinfo {author} {\bibfnamefont {S.-K.}\ \bibnamefont {Son}},
  \bibinfo {author} {\bibfnamefont {T.~R.~M.}\ \bibnamefont {Barends}},
  \bibinfo {author} {\bibfnamefont {T.~A.}\ \bibnamefont {White}}, \bibinfo
  {author} {\bibfnamefont {A.}~\bibnamefont {Barty}}, \bibinfo {author}
  {\bibfnamefont {S.}~\bibnamefont {Botha}}, \bibinfo {author} {\bibfnamefont
  {S.}~\bibnamefont {Boutet}}, \bibinfo {author} {\bibfnamefont
  {C.}~\bibnamefont {Caleman}}, \bibinfo {author} {\bibfnamefont {R.~B.}\
  \bibnamefont {Doak}}, \bibinfo {author} {\bibfnamefont {M.~H.}\ \bibnamefont
  {Nanao}}, \bibinfo {author} {\bibfnamefont {K.}~\bibnamefont {Nass}},
  \bibinfo {author} {\bibfnamefont {R.~L.}\ \bibnamefont {Schoeman}}, \bibinfo
  {author} {\bibfnamefont {N.}~\bibnamefont {Timneanu}}, \bibinfo {author}
  {\bibfnamefont {R.}~\bibnamefont {Santra}}, \bibinfo {author} {\bibfnamefont
  {I.}~\bibnamefont {Schlichting}}, \ and\ \bibinfo {author} {\bibfnamefont
  {H.~N.}\ \bibnamefont {Chapman}},\ }\href {\doibase
  10.1107/S2052252515014049} {\bibfield  {journal} {\bibinfo  {journal}
  {IUCrJ}\ }\textbf {\bibinfo {volume} {2}},\ \bibinfo {pages} {627} (\bibinfo
  {year} {2015})}\BibitemShut {NoStop}%
\bibitem [{\citenamefont {Hao}\ \emph {et~al.}(2015)\citenamefont {Hao},
  \citenamefont {Inhester}, \citenamefont {Hanasaki}, \citenamefont {Son},\
  and\ \citenamefont {Santra}}]{Hao15}%
  \BibitemOpen
  \bibfield  {author} {\bibinfo {author} {\bibfnamefont {Y.}~\bibnamefont
  {Hao}}, \bibinfo {author} {\bibfnamefont {L.}~\bibnamefont {Inhester}},
  \bibinfo {author} {\bibfnamefont {K.}~\bibnamefont {Hanasaki}}, \bibinfo
  {author} {\bibfnamefont {S.-K.}\ \bibnamefont {Son}}, \ and\ \bibinfo
  {author} {\bibfnamefont {R.}~\bibnamefont {Santra}},\ }\href {\doibase
  10.1063/1.4919794} {\bibfield  {journal} {\bibinfo  {journal} {Struct. Dyn.}\
  }\textbf {\bibinfo {volume} {2}},\ \bibinfo {pages} {041707} (\bibinfo {year}
  {2015})}\BibitemShut {NoStop}%
\bibitem [{\citenamefont {Son}\ \emph {et~al.}(2011{\natexlab{b}})\citenamefont
  {Son}, \citenamefont {Young},\ and\ \citenamefont {Santra}}]{Son11a}%
  \BibitemOpen
  \bibfield  {author} {\bibinfo {author} {\bibfnamefont {S.-K.}\ \bibnamefont
  {Son}}, \bibinfo {author} {\bibfnamefont {L.}~\bibnamefont {Young}}, \ and\
  \bibinfo {author} {\bibfnamefont {R.}~\bibnamefont {Santra}},\ }\href
  {\doibase 10.1103/PhysRevA.83.033402} {\bibfield  {journal} {\bibinfo
  {journal} {Phys. Rev. A}\ }\textbf {\bibinfo {volume} {83}},\ \bibinfo
  {pages} {033402} (\bibinfo {year} {2011}{\natexlab{b}})}\BibitemShut
  {NoStop}%
\bibitem [{\citenamefont {Brunvoll}\ \emph {et~al.}(1990)\citenamefont
  {Brunvoll}, \citenamefont {Samdal}, \citenamefont {Thomassen}, \citenamefont
  {Vilkov},\ and\ \citenamefont {Volden}}]{Brunvoll90}%
  \BibitemOpen
  \bibfield  {author} {\bibinfo {author} {\bibfnamefont {J.}~\bibnamefont
  {Brunvoll}}, \bibinfo {author} {\bibfnamefont {S.}~\bibnamefont {Samdal}},
  \bibinfo {author} {\bibfnamefont {H.}~\bibnamefont {Thomassen}}, \bibinfo
  {author} {\bibfnamefont {L.~V.}\ \bibnamefont {Vilkov}}, \ and\ \bibinfo
  {author} {\bibfnamefont {H.~V.}\ \bibnamefont {Volden}},\ }\href {\doibase
  10.3891/acta.chem.scand.44-0023} {\bibfield  {journal} {\bibinfo  {journal}
  {Acta Chem. Scand.}\ }\textbf {\bibinfo {volume} {44}},\ \bibinfo {pages}
  {23} (\bibinfo {year} {1990})}\BibitemShut {NoStop}%
\bibitem [{\citenamefont {Johnson}(2015)}]{Johnson15}%
  \BibitemOpen
  \bibfield  {author} {\bibinfo {author} {\bibfnamefont {R.~D.}\ \bibnamefont
  {Johnson}},\ }\href {http://cccbdb.nist.gov/} {\emph {\bibinfo {title} {NIST
  Computational Chemistry Comparison and Benchmark Database}}},\ \bibinfo
  {type} {Release 17b}\ (\bibinfo  {institution} {National Institute of
  Standards and Technology},\ \bibinfo {year} {2015})\BibitemShut {NoStop}%
\bibitem [{\citenamefont {Sch{\"a}fer}\ \emph {et~al.}(2018)\citenamefont
  {Sch{\"a}fer}, \citenamefont {Inhester}, \citenamefont {Son}, \citenamefont
  {Fink},\ and\ \citenamefont {Santra}}]{Schaefer18}%
  \BibitemOpen
  \bibfield  {author} {\bibinfo {author} {\bibfnamefont {J.~M.}\ \bibnamefont
  {Sch{\"a}fer}}, \bibinfo {author} {\bibfnamefont {L.}~\bibnamefont
  {Inhester}}, \bibinfo {author} {\bibfnamefont {S.-K.}\ \bibnamefont {Son}},
  \bibinfo {author} {\bibfnamefont {R.~F.}\ \bibnamefont {Fink}}, \ and\
  \bibinfo {author} {\bibfnamefont {R.}~\bibnamefont {Santra}},\ }\href
  {\doibase 10.1103/PhysRevA.97.053415} {\bibfield  {journal} {\bibinfo
  {journal} {Phys. Rev. A}\ }\textbf {\bibinfo {volume} {97}},\ \bibinfo
  {pages} {053415} (\bibinfo {year} {2018})}\BibitemShut {NoStop}%
\bibitem [{\citenamefont {Kuleff}\ \emph {et~al.}(2005)\citenamefont {Kuleff},
  \citenamefont {Breidbach},\ and\ \citenamefont {Cederbaum}}]{Kuleff05}%
  \BibitemOpen
  \bibfield  {author} {\bibinfo {author} {\bibfnamefont {A.~I.}\ \bibnamefont
  {Kuleff}}, \bibinfo {author} {\bibfnamefont {J.}~\bibnamefont {Breidbach}}, \
  and\ \bibinfo {author} {\bibfnamefont {L.~S.}\ \bibnamefont {Cederbaum}},\
  }\href {\doibase 10.1063/1.1961341} {\bibfield  {journal} {\bibinfo
  {journal} {J. Chem. Phys.}\ }\textbf {\bibinfo {volume} {123}},\ \bibinfo
  {pages} {044111} (\bibinfo {year} {2005})}\BibitemShut {NoStop}%
\bibitem [{\citenamefont {Remacle}\ and\ \citenamefont
  {Levine}(2006)}]{Remacle06}%
  \BibitemOpen
  \bibfield  {author} {\bibinfo {author} {\bibfnamefont {F.}~\bibnamefont
  {Remacle}}\ and\ \bibinfo {author} {\bibfnamefont {R.~D.}\ \bibnamefont
  {Levine}},\ }\href@noop {} {\bibfield  {journal} {\bibinfo  {journal} {Proc.
  Natl. Acad. Sci. U. S. A.}\ }\textbf {\bibinfo {volume} {103}},\ \bibinfo
  {pages} {6793} (\bibinfo {year} {2006})}\BibitemShut {NoStop}%
\bibitem [{\citenamefont {Vinko}\ \emph {et~al.}(2012)\citenamefont {Vinko},
  \citenamefont {Ciricosta}, \citenamefont {Cho}, \citenamefont {Engelhorn},
  \citenamefont {Chung}, \citenamefont {Brown}, \citenamefont {Burian},
  \citenamefont {Chalupsk{\'y}}, \citenamefont {Falcone}, \citenamefont
  {Graves}, \citenamefont {H{\'a}jkov{\'a}}, \citenamefont {Higginbotham},
  \citenamefont {Juha}, \citenamefont {Krzywinski}, \citenamefont {Lee},
  \citenamefont {Messerschmidt}, \citenamefont {Murphy}, \citenamefont {Ping},
  \citenamefont {Scherz}, \citenamefont {Schlotter}, \citenamefont {Toleikis},
  \citenamefont {Turner}, \citenamefont {Vysin}, \citenamefont {Wang},
  \citenamefont {Wu}, \citenamefont {Zastrau}, \citenamefont {Zhu},
  \citenamefont {Lee}, \citenamefont {Heimann}, \citenamefont {Nagler},\ and\
  \citenamefont {Wark}}]{Vinko12}%
  \BibitemOpen
  \bibfield  {author} {\bibinfo {author} {\bibfnamefont {S.~M.}\ \bibnamefont
  {Vinko}}, \bibinfo {author} {\bibfnamefont {O.}~\bibnamefont {Ciricosta}},
  \bibinfo {author} {\bibfnamefont {B.~I.}\ \bibnamefont {Cho}}, \bibinfo
  {author} {\bibfnamefont {K.}~\bibnamefont {Engelhorn}}, \bibinfo {author}
  {\bibfnamefont {H.-K.}\ \bibnamefont {Chung}}, \bibinfo {author}
  {\bibfnamefont {C.~R.~D.}\ \bibnamefont {Brown}}, \bibinfo {author}
  {\bibfnamefont {T.}~\bibnamefont {Burian}}, \bibinfo {author} {\bibfnamefont
  {J.}~\bibnamefont {Chalupsk{\'y}}}, \bibinfo {author} {\bibfnamefont {R.~W.}\
  \bibnamefont {Falcone}}, \bibinfo {author} {\bibfnamefont {C.}~\bibnamefont
  {Graves}}, \bibinfo {author} {\bibfnamefont {V.}~\bibnamefont
  {H{\'a}jkov{\'a}}}, \bibinfo {author} {\bibfnamefont {A.}~\bibnamefont
  {Higginbotham}}, \bibinfo {author} {\bibfnamefont {L.}~\bibnamefont {Juha}},
  \bibinfo {author} {\bibfnamefont {J.}~\bibnamefont {Krzywinski}}, \bibinfo
  {author} {\bibfnamefont {H.~J.}\ \bibnamefont {Lee}}, \bibinfo {author}
  {\bibfnamefont {M.}~\bibnamefont {Messerschmidt}}, \bibinfo {author}
  {\bibfnamefont {C.~D.}\ \bibnamefont {Murphy}}, \bibinfo {author}
  {\bibfnamefont {Y.}~\bibnamefont {Ping}}, \bibinfo {author} {\bibfnamefont
  {A.}~\bibnamefont {Scherz}}, \bibinfo {author} {\bibfnamefont
  {W.}~\bibnamefont {Schlotter}}, \bibinfo {author} {\bibfnamefont
  {S.}~\bibnamefont {Toleikis}}, \bibinfo {author} {\bibfnamefont {J.~J.}\
  \bibnamefont {Turner}}, \bibinfo {author} {\bibfnamefont {L.}~\bibnamefont
  {Vysin}}, \bibinfo {author} {\bibfnamefont {T.}~\bibnamefont {Wang}},
  \bibinfo {author} {\bibfnamefont {B.}~\bibnamefont {Wu}}, \bibinfo {author}
  {\bibfnamefont {U.}~\bibnamefont {Zastrau}}, \bibinfo {author} {\bibfnamefont
  {D.}~\bibnamefont {Zhu}}, \bibinfo {author} {\bibfnamefont {R.~W.}\
  \bibnamefont {Lee}}, \bibinfo {author} {\bibfnamefont {P.~A.}\ \bibnamefont
  {Heimann}}, \bibinfo {author} {\bibfnamefont {B.}~\bibnamefont {Nagler}}, \
  and\ \bibinfo {author} {\bibfnamefont {J.~S.}\ \bibnamefont {Wark}},\
  }\href@noop {} {\bibfield  {journal} {\bibinfo  {journal} {Nature}\ }\textbf
  {\bibinfo {volume} {482}},\ \bibinfo {pages} {59} (\bibinfo {year}
  {2012})}\BibitemShut {NoStop}%
\bibitem [{\citenamefont {Ziaja}\ \emph {et~al.}(2015)\citenamefont {Ziaja},
  \citenamefont {Jurek}, \citenamefont {Medvedev}, \citenamefont {Saxena},
  \citenamefont {Son},\ and\ \citenamefont {Santra}}]{Ziaja15}%
  \BibitemOpen
  \bibfield  {author} {\bibinfo {author} {\bibfnamefont {B.}~\bibnamefont
  {Ziaja}}, \bibinfo {author} {\bibfnamefont {Z.}~\bibnamefont {Jurek}},
  \bibinfo {author} {\bibfnamefont {N.}~\bibnamefont {Medvedev}}, \bibinfo
  {author} {\bibfnamefont {V.}~\bibnamefont {Saxena}}, \bibinfo {author}
  {\bibfnamefont {S.-K.}\ \bibnamefont {Son}}, \ and\ \bibinfo {author}
  {\bibfnamefont {R.}~\bibnamefont {Santra}},\ }\href {\doibase
  10.3390/photonics2010256} {\bibfield  {journal} {\bibinfo  {journal}
  {Photonics}\ }\textbf {\bibinfo {volume} {2}},\ \bibinfo {pages} {256}
  (\bibinfo {year} {2015})}\BibitemShut {NoStop}%
\bibitem [{\citenamefont {Peltz}\ \emph {et~al.}(2012)\citenamefont {Peltz},
  \citenamefont {Varin}, \citenamefont {Brabec},\ and\ \citenamefont
  {Fennel}}]{Peltz12}%
  \BibitemOpen
  \bibfield  {author} {\bibinfo {author} {\bibfnamefont {C.}~\bibnamefont
  {Peltz}}, \bibinfo {author} {\bibfnamefont {C.}~\bibnamefont {Varin}},
  \bibinfo {author} {\bibfnamefont {T.}~\bibnamefont {Brabec}}, \ and\ \bibinfo
  {author} {\bibfnamefont {T.}~\bibnamefont {Fennel}},\ }\href {\doibase
  10.1088/1367-2630/14/6/065011} {\bibfield  {journal} {\bibinfo  {journal}
  {New J. Phys.}\ }\textbf {\bibinfo {volume} {14}},\ \bibinfo {pages} {065011}
  (\bibinfo {year} {2012})}\BibitemShut {NoStop}%
\bibitem [{\citenamefont {Hau-Riege}\ \emph {et~al.}(2013)\citenamefont
  {Hau-Riege}, \citenamefont {Weisheit}, \citenamefont {Castor}, \citenamefont
  {London}, \citenamefont {Scott},\ and\ \citenamefont
  {Richards}}]{Hau-Riege13a}%
  \BibitemOpen
  \bibfield  {author} {\bibinfo {author} {\bibfnamefont {S.~P.}\ \bibnamefont
  {Hau-Riege}}, \bibinfo {author} {\bibfnamefont {J.}~\bibnamefont {Weisheit}},
  \bibinfo {author} {\bibfnamefont {J.~I.}\ \bibnamefont {Castor}}, \bibinfo
  {author} {\bibfnamefont {R.~A.}\ \bibnamefont {London}}, \bibinfo {author}
  {\bibfnamefont {H.}~\bibnamefont {Scott}}, \ and\ \bibinfo {author}
  {\bibfnamefont {D.~F.}\ \bibnamefont {Richards}},\ }\href {\doibase
  10.1088/1367-2630/15/1/015011} {\bibfield  {journal} {\bibinfo  {journal}
  {New J. Phys.}\ }\textbf {\bibinfo {volume} {15}},\ \bibinfo {pages} {015011}
  (\bibinfo {year} {2013})}\BibitemShut {NoStop}%
\bibitem [{\citenamefont {Jurek}\ \emph {et~al.}(2016)\citenamefont {Jurek},
  \citenamefont {Son}, \citenamefont {Ziaja},\ and\ \citenamefont
  {Santra}}]{Jurek16}%
  \BibitemOpen
  \bibfield  {author} {\bibinfo {author} {\bibfnamefont {Z.}~\bibnamefont
  {Jurek}}, \bibinfo {author} {\bibfnamefont {S.-K.}\ \bibnamefont {Son}},
  \bibinfo {author} {\bibfnamefont {B.}~\bibnamefont {Ziaja}}, \ and\ \bibinfo
  {author} {\bibfnamefont {R.}~\bibnamefont {Santra}},\ }\href {\doibase
  10.1107/S1600576716006014} {\bibfield  {journal} {\bibinfo  {journal} {J.
  Appl. Cryst.}\ }\textbf {\bibinfo {volume} {49}},\ \bibinfo {pages} {1048}
  (\bibinfo {year} {2016})}\BibitemShut {NoStop}%
\bibitem [{\citenamefont {Ho}\ and\ \citenamefont {Knight}(2017)}]{Ho17}%
  \BibitemOpen
  \bibfield  {author} {\bibinfo {author} {\bibfnamefont {P.~J.}\ \bibnamefont
  {Ho}}\ and\ \bibinfo {author} {\bibfnamefont {C.}~\bibnamefont {Knight}},\
  }\href {\doibase 10.1088/1361-6455/aa69e6} {\bibfield  {journal} {\bibinfo
  {journal} {J. Phys. B: At. Mol. Opt. Phys.}\ }\textbf {\bibinfo {volume}
  {50}},\ \bibinfo {pages} {104003} (\bibinfo {year} {2017})}\BibitemShut
  {NoStop}%
\end{thebibliography}
%

\end{document}